\documentclass[11pt, a4paper]{article}
\usepackage{graphicx,epsfig,tikz}
\usepackage{amssymb,amsmath,mathdots}
\usepackage{setspace}
\usepackage{stmaryrd}
\usepackage{mathtools}
\usepackage[toc,page]{appendix}
\usepackage{siunitx}

\newcommand{\bb}{\mathcal{B}}

\newcommand{\bigO}[1]{\ensuremath{\mathop{}\mathopen{}O\mathopen{}\left(#1\right)}}
   
\DeclareMathOperator{\tr}{tr}

\newcommand{\be}{\begin{equation}} 
\newcommand{\en}{\end{equation}}

\def\bga#1\ega{\begin{gather}#1\end{gather}} % suggested in technote.tex
\def\bgas#1\egas{\begin{gather*}#1\end{gather*}}

\def\bal#1\eal{\begin{align}#1\end{align}} % suggested in technote.tex
\def\bals#1\eals{\begin{align*}#1\end{align*}}

\newcommand{\ii}{\textrm{i}}

\renewcommand{\vec}[1]{\boldsymbol{#1}}

\def\scalemath{0.76}

\newcommand*{\Scale}[2][4]{\scalebox{#1}{$#2$}}%
\def \ini#1{\overset{\circ}{ #1}} 

\newtheorem{example}{Example}

\usepackage[style= bwl-FU, backend = bibtex, maxcitenames=2,uniquelist=minyear, firstinits=true]{biblatex}
\AtEveryBibitem{\clearfield{number}}
\AtEveryBibitem{\clearfield{doi}}   
\AtEveryBibitem{\clearfield{url}}   
\AtEveryBibitem{\clearfield{issn}}
\AtEveryBibitem{\clearfield{isbn}}

\usepackage{geometry}
\graphicspath{{../Images/}}

\begin{document}
\renewcommand{\thefootnote}{\fnsymbol{footnote}}

\title{\textsc{Initial Stress Symmetry and} \\ \textsc{ Applications in Elasticity }}
\author{
 A. L. Gower\footnotemark[2] \footnotemark[6], P. Ciarletta\footnotemark[4] \footnotemark[5], M. Destrade\footnotemark[2] \footnotemark[3]  
 %\\[12pt]
%$^a$ School of Mathematics, Statistics and Applied Mathematics,\\
%National University of Ireland Galway;\\
%University Road, Galway, Ireland,\\[12pt]
}
\footnotetext[6]{Corresponding author. E-mail: arturgower@gmail.com}
\footnotetext[2]{School of Mathematics, Statistics and Applied Mathematics,
National University of Ireland Galway,
University Road, Galway, Ireland.}
\footnotetext[3]{School of Mechanical \& Materials Engineering,  University College Dublin,
Belfield, Dublin 4, Ireland.}
\footnotetext[4]{
CNRS and Institut Jean le Rond d'Alembert, UMR
7190,
 Universit$\rm \acute{e}$ Paris 6,
  4 place Jussieu case 162, 75005 Paris, France.} 
\footnotetext[5]{
MOX - Politecnico di Milano and Fondazione CEN, piazza
Leonardo da Vinci 32, 20133 Milano, Italy.}

\date{\today}
\maketitle

\begin{abstract}
An initial stress within a solid can arise to support external loads or from processes such as thermal expansion in inert matter or growth and remodelling in living materials.  
For this reason it is useful to develop a mechanical framework of initially stressed solids irrespective of how this stress formed.   
An ideal way to do this is to write the free energy density $\Psi= \Psi(\vec F, \vec \tau)$ in terms of initial stress $\vec \tau$ and the elastic deformation gradient $\vec F$.
 %is ideal for in-vivo biological tissues.  
 In this paper we present a new constitutive condition for initially stressed materials, which we call the initial stress symmetry (ISS). We focus on two consequences of this symmetry. First we examine how ISS restricts the free energy density $\Psi = \Psi (\vec F, \vec \tau) $ and present two examples of $\Psi (\vec F, \vec \tau)$ that satisfy ISS. Second we show that the initial stress can be derived from the Cauchy stress and the elastic deformation gradient. To illustrate we take an example from biomechanics and calculate the optimal Cauchy stress within an artery subjected to internal pressure. We then use ISS to derive the optimal target residual stress for the material to achieve after remodelling.
\end{abstract}

\noindent
{\textit{Keywords:} residual stress, initial stress, biomechanics, elasticity, constitutive equations}

%%%%%%%%%%%%

\section{Introduction}

%%%%%%%%%%%%%

When all loads are removed a body can still hold a significant amount of internal stress, called the \emph{residual stress}. In manufacturing, residual stress has long been noted to be detrimental to, or enhance, the performance of a material.
%~\parencite{withers2001residual}. 
For biological tissues, residual stress is used to self-regulate stress and strain, and ultimately preserve ideal mechanical conditions for the tissue~\parencite{fung1991residual,holzapfel2003biomechanics}. In geophysics, due to gravity the Earth has developed high initial stress within, which greatly influence the propagation of elastic waves. 

Here we use the term \emph{initial stress} to broadly mean the internal stress of some reference configuration, irrespective of how the stress was formed or the boundary conditions. In this sense residual stress is a form of initial stress.

The initial stress felt by any region of a material is due to the push and pull of the surrounding regions. If any region were to be cut out from the material, the stress on its newly formed boundary would be zero, thus reducing the potential energy in the bulk. Based on this concept Hoger developed constitutive laws for residually stressed materials, see for example~\cite{hoger1986determination, hoger1993elasticity, johnson1995use}. 
Hoger showed that by taking this idea to its limit, and cutting the material into possibly an infinite number of disconnected regions, the material may be relieved of all of its internal stress in a configuration called the \emph{virtual stress-free state}. From that configuration, a hyperelastic energy can be defined as a function of the strain from the virtual state to the current configuration.  

Though the use of this virtual state is technically sound, it leads to challenging calculations even for simple deformations and rarely yields analytic results, unless great simplifications are assumed.  
%For example, the often-used ``open angle method'' cite (and elaborate). 
Moreover, the experimentally identification of the virtual state requires cutting the material, which is not always suitable, especially for living organisms. However, using a virtual stress-free reference is routinely seen as the only viable alternative, quoting \cite{chuong1986residual}: ``To characterize the arterial wall or any other biological soft tissue, we need a stress-free state''. Conversely, We believe that by developing tools to work directly with initially stressed reference configurations, without the need of a stress-free state, will be very useful, specially in biomechanics. 

An ideal way to account for the initial stress would be to have a free energy density function $\Psi= \Psi(\vec F, \vec \tau)$ written explicitly in terms of the deformation gradient $\vec F$ and the initial stress $\vec \tau$, without any a priori restrictions.  For the development of constitutive laws there is no need to distinguish between residual stresses and initial stresses, a view which is shared with~\cite{merodio2013influence}. The initial stress $\vec \tau$ could then be determined from elastic wave speeds~\parencite{shams2014rayleigh,destrade2012stress,man1987towards} or by solving the linear equations of momentum balance. 
%The effect of the residual stress would then be expressed through the Cauchy stress Eq.~\eqref{eqn:CauchyStressGeneral}.
 %rather than nonlinear equations often appear when calculating compatible configurations for virtual states (<- is this true Pasquale?).

\cite{shams2011initial} and~\cite{shams2010wave} worked towards a general framework for initially stressed solids, while others have investigated the mechanics for some examples of $\Psi(\vec F, \vec \tau)$~\parencite{merodio2013influence,merodio2015extension}. For a more complex geometry, \cite{wang2014modified} found that including one residual stress invariant in the free energy density was a simple way to model the effects of residual stress on the myocardium.   
However, in general a major obstacle still remains: how to write $\Psi(\vec F, \vec \tau)$ in terms of the combined invariants of $\vec F$ and $\vec \tau$? If the only source of anisotropy is due to the residual stress, then the free energy still depends on ten independent invariants.
 %see equations~\eqref{eqns:Is},\eqref{eqns:Itau} and \eqref{eqns:Icombined}.

\cite{johnson1995use} developed representations for the Cauchy stress response $\vec \sigma =\hat {\vec \varsigma}(\vec F ,\vec \tau )$ in terms of $\vec F$ and $\vec \tau$ by assuming a stress free virtual state and numerically inverting the residual stress-strain equation. However, this approach often requires  solving numerical nonlinear implicit equations. \cite{johnson1995use,johnson1998use} exemplified this approach for a material with virtual state composed by a Mooney-Rivlin strain energy.

%cooling down metals during fabrication introduces residual stress, due to the mismatch of thermal expansion~\cite{arsenault1987thermal},

In this work, we introduce a new constitutive requirement on $\Psi(\vec F ,\vec \tau )$ called the Initial Stress Symmetry (ISS). ISS restricts the constitutive form of the stress $\vec \sigma =\hat {\vec \varsigma}(\vec F ,\vec \tau )$ by providing a constitutive equation for the initial stress $\vec \tau =\hat {\vec \varsigma}(\vec F^{-1} ,\vec \sigma )$ that must hold for every $\vec F$ and $\vec \tau$. To our knowledge this symmetry has never been discussed before.  

The ISS also helps answer an important question: how much can the Cauchy stress be altered by adjusting the initial stress? From a modelling perspective, initial stress has been used to make the material more or less compliant~\parencite{johnson1998use}, to control the Poynting effect~\parencite{merodio2013influence} or to maintain an ideal internal stress~\parencite{fung1991residual}. Given a $\Psi(\vec F ,\vec \tau )$ that satisfies ISS, then $\vec \tau =\hat {\vec \varsigma}(\vec F^{-1} ,\vec \sigma )$ suggests that for any choice of $\vec \sigma$, there will exist an initial stress $\vec \tau$ that supports $\vec \sigma$. 

%As an example we use this principle to  determine the residual stress supports the ideal  
The basic equations for an elastic material subject to initial stress are summarized in Section~\ref{sec:Hyperelastic}, and then the initial stress symmetry is first presented in Section~\ref{sec:introISS}. ISS is satisfied automatically if the initial stress is due to an elastic deformation of a stress free configuration, which we demonstrate in Appendix~\ref{app:VirtualConfig}. For this reason we develop in Section~\ref{sec:SimpleEnergy} an example for $\Psi(\vec F ,\vec \tau )$, that satisfies ISS, by deforming an incompressbile neo-Hookean material from a stress free virtual state. 

%An excellent example of $\Psi(\vec F, \vec \tau)$ that satisfies ISS results from assuming that the initial/residual stress is due to deforming an incompressible neo-Hookean solid from a virtual stress configuration. We reach a closed form solution for this example, see equation.

In Section~\ref{sec:ISS} we express ISS, in all generality, as nine scalar equations for an incompressible material, written in terms of $\Psi$; two undetermined scalars $p$ and $p_\tau$; and the invariants of $\vec F$ and $\vec \tau$. This form of ISS makes it easier to select representations for $\Psi(\vec F ,\vec \tau )$. For example, in Section~\ref{sec:independentJ_3&J4} we show, with a minor adjustament to the equations, how to use the scalar equations of ISS to deduce an example of $\Psi(\vec F ,\vec \tau )$ for a compressibly material. 
 
%Finally, even when a virtual stress-free state is at hand, finding a compatible deformation that maps it into a balanced unloaded configuration generally involves solving complicated coupled nonlinear equations cite (\emph{does that make sense, I mean nonlinear elasticity leads to nonlinear equations.}), a process which very rarely yields analytical results. 

It is commonly thought that arteries attempt to maintain a homogeneous stress gradient within their walls~\parencite{taber1996theoretical}.
%We exemplify this by using the residual stress to maintain an optimal Cauchy stress for a simplified arterial wall in Section~\ref{sec:HomogeneousStress}.  
In Section~\ref{sec:HomogeneousStress} we calculate this optimal Cauchy stress for a simplified arterial wall, and then show how by using ISS we can calculate the residual stress that exactly supports this optimal Cauchy stress in Section~\ref{sec:Inflation}. We finally compare these results against the commonly used opening-angle method~\parencite{chuong1986residual}. 
%and show that the two methods give similar results for small values of $P/\mu$.

\section{Initially stressed elastic materials}
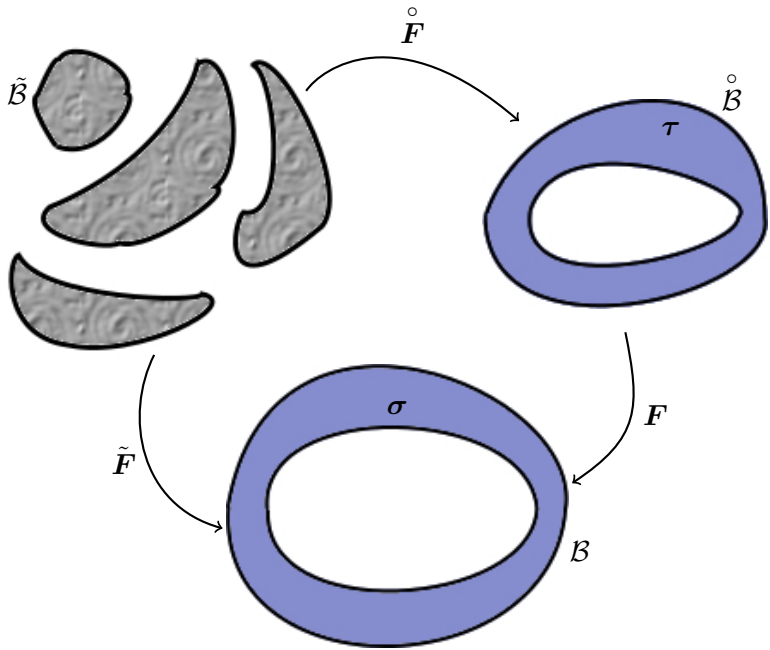
\begin{figure}[ht!]
\centering{
\begin{tikzpicture} 
 	\draw (0,0) node {\includegraphics[width=0.3\textwidth ]{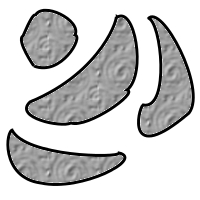} }; 
 	\draw (6,0) node {\includegraphics[width=0.25\textwidth ]{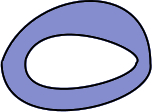} } ;
 	\draw (3,- 4) node {\includegraphics[width=0.3\textwidth ]{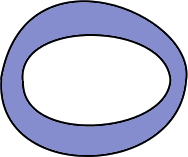} } ;
 	
 	\draw(6.6,1.) node[thick]{$ \vec \tau$};
 	\draw(3.,-2.7) node{$ \vec \sigma$};
 	
 	\draw(-2.,1.5) node{$\tilde {\bb}$};
 	\draw(7.4,1.5) node{$ \ini {\bb}$};
 	\draw(5.4,-4.6) node{$ {\bb}$};
 	
 	\draw [->] (1.8,1.5) .. controls (2.,1.8) and (3,2.5) .. (4.6,1.1)[thick]  ;
 	\draw [->] (6.,1.3-3) .. controls (6.2,1.3-3-1.) and (6.2,1.3-4.4) .. (5.3,1.3-5)[thick] ;
 	\draw [<-] (0.7,-4.3) .. controls (-.4,-4) and (-0.6,-2.8) .. (-.2,-2)[thick]  ;
	
 	\draw(3.2,2.) node[above]{$\ini {\vec F}$};
 	\draw(6.1,-2.8) node[right]{$ \vec F$};
 	\draw(-.3,-3.4) node[left]{$ \tilde {\vec F}$};
\end{tikzpicture}
}
\caption{
%The possible configurations and kinematics for an initially stressed elastic body. 
$\bb$ is the current configuration with internal stress $\vec \sigma$, while $\ini \bb$ is a reference configuration with internal stress $\vec \tau$. The virtual stress-free state $\tilde {\bb}$ is a collection of configurations where the body is stress-free.}
\label{fig:Configurations}
\end{figure}

%%%%%%%%%%%%%%%%%%%%%%%%%%%%%%%%%%%
\label{sec:Hyperelastic}
 A common approach to model the effect of \emph{residual stress} is to consider a \emph{virtual stress free configuration} $\tilde{\bb}$, from which the material is deformed and ``glued'' together to produce a residually stressed equilibrium state $\ini \bb$. See Figure~\ref{fig:Configurations} for a diagram of all the configurations. An elastic stored energy density $\Psi$ can then be defined as a function of the \emph{deformation gradient} $\tilde {\vec F}$ from $\tilde {\bb}$ to the current configuration $\bb$ so that $\Psi = \Psi( \tilde {\vec F})$. 

In this paper we want to write the free energy density $\Psi$ as a function of the initial stress $\vec \tau$ and of the deformation gradient ${\vec F}:\ini{\bb} \to \bb$, so that $\Psi=\Psi(\vec F, \vec \tau)$, and $\ini \bb$ will not necessarily be an unloaded configuration. We feel that it is natural to consider that initial stress contributes to the potential energy stored by a material. One extreme example is the wapa tree, which has been know to burst open once cut, possibly causing injury, due to its immense level of residual stress~\parencite{Wapas1988}. 
%Once an expression for $\Psi$ in terms of $\vec \tau$ and $\vec F$ has been determined, $\vec \tau$ can be calculated from the equations of equilibrium and the boundary conditions in the unloaded configuration, both of which are linear equations. It is even conceivable that the residual stress $\vec \tau$ can be measured from experiments using acoustic waves~\parencite{shams2011initial} or standard tensile tests.

%To express $\Psi$ as a general function of the two independent variables $\vec \tau$ and $\vec F$ we let $\Psi=\Psi(\vec F, \vec \tau)$. 
We assume that $\vec F$ is a purely elastic deformation, but we do not make any assumptions about the origins of the initial stress $\vec \tau$, except that $\vec \tau$ affects the stored energy density $\Psi$. Assuming that the body is incompressible, i.e. $J= \det {\vec F}= 1$ at all times, the Cauchy stress tensor ${\boldsymbol \sigma}$ reads \parencite{Guillou2006Residual}:
\be
{\vec \sigma}= \vec F \frac{\partial \Psi}{\partial {\vec F}}({\vec F},
{\vec \tau})-p \vec I,
\label{eqn:CauchyStressGeneral}
\en
where $p$ is the Lagrange multiplier associated with the constraint of incompressibility, $\vec{I}$ is the identity matrix  and $\vec \tau$ is the initial stress. If the material is compressible then $p$ is replaced by $- 2 I_3 \partial \Psi/ \partial I_3$. Note we have and will omit the possible dependence of $\Psi$ on the position $X \in \ini \bb$ for the sake of simplicity.
 For the body in the configuration $\ini \bb$ we have that $\vec F = \vec I$ and $\vec \sigma = \vec \tau$; thus we require that
\be 
\vec \tau= \frac{\partial \Psi}{\partial {\vec F}}({\vec I}, \vec \tau)-\ini p {\vec I}, 
\label{eqn:StressCompatibility}
\en
where $\ini p$ is the value of $p$ when $\vec F = \vec I$. We call the above equation the \emph{residual stress compatibility}.

The presence of residual stress generally leads to an anisotropy response of the material in reference to $\ini \bb$. Here we assume no other source of intrinsic anisotropic so that $\Psi$ can be written as a function of all the independent invariants generated by $\vec \tau$ and $\vec C = \vec F^T \vec F$, the right Cauchy-Green deformation tensor. Following \citeauthor{shams2011initial}(2011), we take the following complete set of 10 independent invariants
\bal
& I_1= \tr {\vec C}, \quad I_2=\frac{1}{2}[(I_1^2 -\tr({\vec C }^2)],
\quad  I_3= \det {\vec C} 
\label{eqns:Is},
\\
& I_{\tau_1}= \tr \vec \tau, \quad  I_{\tau_2}=\frac{1}{2}[(I_{\tau_1}^2 -\tr(\vec \tau^2)],
\quad  I_{\tau_3}= \det \vec \tau 
\label{eqns:Itau},
\\
& J_1= \tr (\vec \tau{\vec C}), \quad J_2= \tr (\vec \tau {\vec C}^2), \quad J_3= \tr ({\vec
\tau}^2 {\vec C}), \quad J_4= \tr (\vec \tau^2 {\vec C}^2).
\label{eqns:Icombined}
\eal
The Cauchy stress Eq.~\eqref{eqn:CauchyStressGeneral} can then be written as~\parencite{shams2011initial} 
\begin{multline}
\vec \sigma = 2 \Psi_{I_1} {\vec B}+2 \Psi_{I_2} (I_1{\vec B}-{\vec B}^2)-p {\vec I}+2 \Psi_{J_1}{\vec F \vec \tau \vec F^T} +2 \Psi_{J_2} \vec F({ \vec \tau }{\vec C}+{\vec C}{\vec \tau })\vec F^T
\\ 
 +2 \Psi_{J_3} \vec F \vec \tau^2 \vec F^T+ 2\Psi_{J_4}\vec F({ \vec \tau^2 \vec C}+{\vec C \vec \tau^2  })\vec F^T
\label{eqn:CauchyInvariants}
\end{multline}
where $\vec B = \vec F \vec F^T$, and $ \Psi_{I_1},\Psi_{I_2}, \Psi_{J_1}, \Psi_{J_2},\Psi_{J_3},\Psi_{J_4}$ are the partial derivatives of $\Psi$ with respect to $I_1, I_2, J_1, J_2,J_3,J_4$ respectively. There are no partial derivative of $\Psi$ with respect to $I_{\tau_1}, \, I_{\tau_2}$ and $I_{\tau_3}$ appearing in~\eqref{eqn:CauchyInvariants} because $\vec \tau$ does not depend on $\vec F$. 

The residual stress compatibility Eq.~\eqref{eqn:StressCompatibility} becomes
\be
2\frac{\partial \Psi}{\partial I_1}+4\frac{\partial \Psi}{\partial I_2}-\ini p=0,
 \quad 2\frac{\partial \Psi}{\partial J_1}+4\frac{\partial \Psi}{\partial J_2}=1,
  \quad \frac{\partial \Psi}{\partial J_3}+2\frac{\partial \Psi}{\partial J_4}=0. 
\label{nc}
\en

Another important physical restriction that can be imposed is the \emph{strong-ellipticity} condition, which is satisfied when the fourth-order tensor  
\[
\mathcal A_{0piqj} = J^{-1} F_{p \alpha}F_{q \beta} \frac{\partial^2 \Psi}{\partial F_{i \alpha} \partial F_{j \beta}  },
\]
satisfies
\be
\label{cond:SE}
\mathcal A_{0piqj} n_p n_q m_i m_j >0 \quad \text{for every} \; \; \vec n, \vec m \in \mathbb R^3, 
\en
for compressible materials, while for incompressible materials the above need only hold for $\vec n \cdot \vec m =0$.    
Imposing SE implies that plane waves may propagate in every direction with a real valued speed~\parencite{truesdell1966existence}, and other physically expected behaviour~\parencite{Walton2003}. For a representation of $\mathcal A_{0piqj}$ in terms of the invariants of $\vec \tau$ and $\vec F$ see~\cite{shams2011initial}.
 
The issue we address now is how to write $\Psi$ explicitly in terms of the invariants~\eqref{eqns:Is}, \eqref{eqns:Itau} and~\eqref{eqns:Icombined}?  We advocate three criteria. The free energy density $\Psi$ should satify the initial stress compatibility~\eqref{nc}. Second, it should satisfy strong-ellipticity~\eqref{cond:SE} for all deformations in which the material is expected to be stable~\parencite{merodio2003instabilities}, and the third criterion we call the \emph{initial stress symmetry} (ISS). 

\subsection{Initial Stress Symmetry}
 \label{sec:introISS}
  For convenience let the response function $\hat{\vec \varsigma}$ be denoted by 
 \be
  \hat {\vec \varsigma }(\vec F_1, \vec \sigma_2, p_1) \coloneqq \vec F_1 \frac{\partial \Psi}{\partial \vec F}(\vec F_1, \vec \sigma_2) -\vec I p_1,
 \en
 for every $\vec F_1$ and $\vec \sigma_2$, where the argument on the right hand side is evaluated by taking the partial derivative of $\Psi(\vec F, \cdot)$ with respect to $\vec F$. The scalar $p_1$ is undetermined if the material is incompressible and $p_1 = - 2 I_3 \Psi_{I_3}$ with $\vec F$ replaced with $\vec F_1$ if the material is compressible. 
 
 The ISS states that $\hat{\vec \varsigma}$ has no preferred reference configuration. 
 %In contract to standard elasticity, the reference state does not need to be a stress-free state. 
 Refering to Figure~\ref{fig:Configurations}, if we take $\ini \bb$ as  the reference configuration, then the Cauchy stress becomes 
 \be
 \label{eqn:CauchyConstitutive}
 \vec \sigma = \hat {\vec \varsigma }(\vec F, \vec \tau, p).
 \en
 However, we can also take $\bb$ as the reference configuration and $\ini\bb$ as the current configuration and therefore express the initial stress as 
\be
 \label{eqn:ResidualISS}
\vec \tau = \hat {\vec \varsigma }(\vec F^{-1}, \vec \sigma, p_\tau),
\en
for some scalar $p_\tau$. For a compressible material $p_\tau = - 2 I_3 \Psi_{I_3}$ with $\vec F$ replace with $\vec F^{-1}$.  
%However, the boundary conditions in $\ini \bb$ and $\bb$ do not matter; the boundary conditions for $\vec \tau$ do not need to be unloaded. 
%For what follows the boundary conditions on $\ini \bb$ and $\bb$ together with Eqs.~\eqref{eqn:CauchyConstitutive} and \eqref{eqn:ResidualISS} should uniquely specify $\vec \sigma$ and $\vec \tau$, but otherwise are arbitrary. 
In a more precise form, ISS can be stated as
 %\be
 %\label{cond:ISS}
 %\vec \sigma_1 = \hat {\vec \varsigma}(\vec F_1,\vec \sigma_2, p_1) \quad \text{and} \quad \vec \sigma_2 = \hat {\vec \varsigma}(\vec F^{-1}_1,\vec \sigma_1, p_2)
 %\en  
 \be
 \label{cond:ISS}
 \vec \sigma = \hat {\vec \varsigma}(\vec F,\vec \tau, p) \quad \text{and} \quad \vec \tau = \hat {\vec \varsigma}(\vec F^{-1},\vec \sigma, p_\tau),
 \en
 for every $\vec F$ and $\vec \tau$, such that $\vec \tau = \vec \tau^T$ and $\det \vec F =1$ for incompressible materials, and  $p$ and $p_\tau$ are respectively given by $p = \hat p (\vec \sigma, \vec F, \vec \tau)$ and $p_\tau = \hat p (\vec \tau, \vec F^{-1}, \vec \sigma)$ for some scalar function $\hat p$. The boundary conditions can determine $p$ through its dependence on $\vec \sigma$, and analogously for $p_\tau$. The ISS agrees with the initial stress compatibility, i.e. the condition~\eqref{eqn:StressCompatibility}, when we have $ \hat{\vec \varsigma} (\vec I , \vec \tau, p)= \vec \tau $ for every $\vec \tau$.  
 
 %, which may put restrictions on $p_1$.  
 %To demonstrate, take $\vec F = \vec I$ then $\vec \sigma_1 = \hat {\vec \varsigma}(\vec I, \vec \sigma_2,p_1)  $ and $\vec \sigma_2 = \hat {\vec \varsigma}(\vec I, \vec \sigma_1,p_2)$ for every $\vec \sigma_2 = \vec \sigma_2^T $, which implies that  $\hat {\vec \varsigma}(\vec I, \vec \sigma_2) = \pm \vec \sigma_2$.
 Another way to view the ISS symmetry is to assume that $\vec \tau$ and $\vec \sigma$ are due to the elastic deformation of a virtual stress-free state, which we demonstrate in Appendix~\ref{app:VirtualConfig}. Hoger emphasized many times that using the virtual stress-free state does not restrict how the residual stress was formed. The same can be said about ISS; only $\vec F$ is an elastic deformation. Moreover, the ISS is not restricted to elasticity and should hold for other constitutive equations such as those encountered in viscoelasticity and plasticity.   
  
One practical outcome from the ISS is to restrict the possible constitutive choices for $\hat{\vec \varsigma}$. 
In Section~\ref{sec:independentJ_3&J4} we show that $\Psi = \frac{1}{2} \mu (I_1 -3) + \frac{1}{2}(J_1 - I_{\tau_1})$ proposed by \cite{merodio2013influence} does not satisfy ISS, and  
we also use ISS to deduce an expression for $\Psi(\vec F, \vec \tau)$ for compressible materials. 
 %if it to satisfy ISS, then the undetermined lagrange multipliers have to be fixed as $p =\mu$ and $p_\tau =\mu$. Having both $p$ and $p_\tau$ determined means that this model can  only be used for specific boundary conditions.

 A second practical feature arising from ISS is that we can write the residual stress as a function of the Cauchy stress~\eqref{eqn:ResidualISS}. This proves useful when $\vec \sigma$ is known a priori, as it will be discussed in Section~\ref{sec:MinimalStressGrad} where we determine $\vec \sigma$ from a homeostasis principal and then use Eq.~\eqref{eqn:ResidualISS} to derive~$\vec \tau$.
%useful when we wish to first choose $\vec \sigma$, such as in Section~\ref{sec:MinimalStressGrad} where we find $\vec \sigma$ with the minimal spatial gradient and then use Eq.~\eqref{eqn:ResidualISS} to solve for $\vec \tau$. 
  The alternative of choosing the residual stress $\vec \tau$ first can be far more complicated. 

Before further developing the implications of ISS in Section~\ref{sec:ISS}, we will introduce below an example of stored energy $\Psi(\vec F, \vec \tau)$ that satisfies ISS, stress compatibility~\eqref{eqn:StressCompatibility} and strong-ellipticity~\eqref{cond:SE}. 

\section{Initially stressed neo-Hookean material}
\label{sec:SimpleEnergy}

Here we derive a simple constitutive equation for an initially stressed body
 %by considering an incompressible neo-Hookean elastic behaviour.
by assuming that both $\vec \tau$ and $\vec \sigma$ arise from deforming an incompressible neo-Hookean material. 
The result will be an explicit representation of $\Psi$ in terms of $\vec F$ and $\vec \tau$, given by Eq.~\eqref{eqn:PsiInTau} below, that automatically satisfies ISS, the stress compatibility~\eqref{nc} and strong-ellipticity~\eqref{cond:SE}. Both ISS and stress compatibility hold because $\vec \tau$ arises due to an elastic deformation from a stress-free state, see Appendix~\ref{app:VirtualConfig}, and strong-ellipticity is satisfied because the material is a neo-Hookean solid~\parencite{ogden1997}.
  %begin by developing a simple example of the free-energy $\Psi$ written in terms of the invariants of $\vec \tau$ and $\vec F$. For this example we make use of a virtual stress-free configuration. The advantages that we gain from doing so is that the resulting energy $\Psi$ always satisfies the initial stress compatibilty~\eqref{eqn:StressCompatibility}, strong-ellipticity (SE)~\eqref{cond:SE} and the initial stress symmetry (ISS)~\eqref{cond:ISS}. 
 
Refering to Figure~\ref{fig:Configurations}, as the material is stress-free in $\tilde \bb$, we have 
\be
\label{eqn:PsineoHook}
\Psi  = \frac{\mu}{2}( \tr  \tilde {\vec C} -3),
\en
where $\mu >0$ is the constant shear modulus and $\tilde {\vec C} = \tilde {\vec F}^T \tilde {\vec F}$. The Cauchy stress~\eqref{eqn:CauchyStressGeneral} then becomes
\be
\label{eqn:CauchyNeoHook}
\vec \sigma = \mu \tilde {\vec F} \tilde {\vec F}^T - p \vec I.
\en 
%Because the stress in $\ini \bb$ is also due to deforming a neo-Hookean material we can write~\eqref{eqn:PsineoHook} as a function of the invariants of $\vec \tau$ and $\vec F$. 
%The Cauchy stress will then be given by~\eqref{eqn:CauchyStressGeneral} or~\eqref{eqn:CauchyInvariants}. 
We can rewrite $\tr \tilde{\vec C}$  by substituting  $\tilde {\vec F} = {\vec F} \ini {\vec F}$ and using the properties of the trace
\[
\tr \tilde{\vec C} = \tr (\tilde {\vec F}^T \tilde {\vec F} ) = \tr (\ini {\vec F^T} \vec F^T  \vec F \ini {\vec F}  ) = \tr (\ini {\vec F} \ini {\vec F^T} \vec F^T  \vec F   ) =  \tr ( \ini {\vec B} \vec C ),
\] 
where $\ini {\vec B} = \ini{\vec F} \ini{\vec F^T} $ and $\vec C = {\vec F}^T {\vec F}$, which leads to
\be
\label{eqn:NeoHookPsi}
\Psi  = \frac{\mu}{2}\left [ \tr ( \ini {\vec B} \vec C ) -3\right ].
\en
 We can write $\ini {\vec B}$ in terms of $\vec \tau$ by evaluating Eq.~\eqref{eqn:CauchyNeoHook} at $\vec F = \vec I$, $\vec \sigma = \vec \tau$, $p = \ini p$ and rearranging the term as follows
\bga
\label{eqn:tau}
  \mu \ini {\vec B} = \vec \tau +\ini p \vec I .
\ega
%which is equivalent to Eq.~\eqref{eqn:StressCompatibility} with $\Psi(\vec I , \vec \tau) = \mu/2 \tr \ini {\vec C} $. %which is analogous to Eq.~\eqref{eqn:SigmaVirtual} with $\Psi = \mu \tr \tilde{ \vec C} /2$.
 %Rearranging the above and taking the trace on either side we reach 
 %\[
 %\tr \ini {\vec B} = \mu^{-1}(I_{\tau1} + 3  \ini p). 
 %\]
 To write $\ini p$ as a function of $\vec \tau$ we require that $\ini {\vec F}$ be isochoric, so that $\det (  \mu \ini {\vec B}) =  \det (\vec \tau + \ini p \vec I) =  \mu^3$, which results in a cubic equation for $\ini p$
\be
   \ini p^3 + \ini p^2 I_{\tau_1} + \ini p I_{\tau_2} +I_{\tau_3} -  \mu^3 =0.
   \label{eqn:inipCubic}
\en The real roots for $\ini p$ are given by,
 %if $T_1 >0$ then
\be
\ini p =
\begin{dcases}
        \frac{1}{3} \left [ T_3+\frac{ T_1}{ T_3 } - I_{\tau_1} \right ] , &  \; \; T_2 \leq T_1^{3/2}   , \\
          \frac{1}{3} \left [ c_1 T_3+  c_1^* \frac{ T_1}{ T_3 } - I_{\tau_1} \right ] , &  \; \;  -T_1^{3/2} \leq  T_2    , \\
          \frac{1}{3} \left [ c_1^* T_3+  c_1 \frac{ T_1}{ T_3 } - I_{\tau_1} \right ] , &  \; \;  -T_1^{3/2} \leq  T_2 \leq T_1^{3/2} , 
   %\mu - \frac{I_{\tau_1}}{3} , &  \; \; T_1=0,
\end{dcases}
\label{eqn:inip}
\en
while if $T_1 =0$ then $\ini p = \mu - I_{\tau_1}/3$,  where
\begin{align}
\label{eqn:ptauTs} 
  T_1=&  I_{\tau_1}^2 -3 I_{\tau_2}, \quad 
   T_2 =  I_{\tau_1}^3   - \frac{9}{2} I_{\tau_1}  I_{\tau_2} + \frac{27}{2} ( I
   _{\tau_3} -\mu^3), \\ 
  T_3 =&  \sqrt[3]{\sqrt{T_2^2-  T_1^3} -  T_2  }, \quad
  c_1=  -\frac{1}{2} +\frac{\sqrt{3}}{2} \ii.
\end{align}  
In terms of the eigenvalues  $\tau_1$, $\tau_2$ and $\tau_3$ of $\vec \tau$, we can write
\begin{align} 
   &T_2 = -\frac{27}{2}\mu^3 +\frac{1}{2} \left[ (\tau_1-\tau_2)- (\tau_3-\tau_1) \right ] \left[  (\tau_2-\tau_3)- (\tau_1-\tau_2) \right ] \left[ (\tau_3-\tau_1)- (\tau_2-\tau_3) \right ] , \notag \\ 
  &T_1= \frac{1}{2}(\tau_1 -\tau_2)^2+ \frac{1}{2}(\tau_1 -\tau_3)^2+ \frac{1}{2}(\tau_2 -\tau_3)^2 , 
  \label{eqn:ptauTs2}
\end{align}
so that when $T_1 =0$ we have that $\tau_1=\tau_2=\tau_3$. From the above we see that $T_1\geq0$  and $T_1^{3/2} \geq  |T_2 + \mu^3 27/2|$ for any\footnote{http://math.stackexchange.com/questions/1255883/prove-that-a2b2c2-geq-2a-b2b-c2a-c21-3/1255907\#1255907} $\tau_1, \tau_2, \tau_3 \in \mathbb R$. So if $T_2< 0$ then clearly $T_2 < T_1^{3/2}$, else if $T_2\geq 0$ then $T_2< |T_2 + \mu^3 27/2| \leq  T_1^{3/2}$. Therefore $T_2 < T_1^{3/2}$, and so the condition for the first case for $\ini p$ in Eq.~\eqref{eqn:inip} is always satisfied. We discard the second and third case for $\ini p$  
because we expect $\Psi$, and therefore $\ini p$, to be continuous for every $\vec \tau \in \mathbb R^3$. When $\tau_1 = \tau_2=\tau_3$ the only viable solution for $\ini p$ is~\eqref{eqn:inip}${}_1$. If $\vec \tau$ moves into a region where~\eqref{eqn:inip}${}_2$ or \eqref{eqn:inip}${}_3$ becomes real, then $\ini p$ can not change from~\eqref{eqn:inip}${}_1$ to~\eqref{eqn:inip}${}_2$ or \eqref{eqn:inip}${}_3$ because it can be shown that \eqref{eqn:inip}${}_1$ does not equal \eqref{eqn:inip}${}_2$ or \eqref{eqn:inip}${}_3$ for any $\vec \tau \in \mathbb R^3$. 

%That is, when $\vec \tau = \vec 0$ the solution for $\ini p$ should be positive and real, this is because $\ini p \vec I = \mu \ini {\vec B} $ when $\vec \tau = \vec 0$, see Eq.\eqref{eqn:tau}, $\mu>0$ and $\ini {\vec B}$ should be positive definite so that $\Psi$ is positive and convex.   

 %The second and third case for $\ini p$ are also possible solutions to Eq.~\eqref{eqn:inipCubic} when additionally $-T_1^{3/2} \leq  T_2$. When this occurs, $\ini p$ becomes multivalued, which means there are three distinct virtual stress-free configurations that lead to the same initial stress $\vec \tau$. 
%Using Eqs.~\eqref{eqn:ptauTs2} it can be shown that if $\mu > 0.68 T_1^{1/2}$ then $-T_1^{3/2} >  T_2$, and $\ini p$ would have a unique real solution.  

%since $3 I_{\tau_2}\leq I_{\tau_1}^2$, which can be shown by writing $I_{\tau_1} = \tau_1+\tau_2+\tau_3$ and $I_{\tau_2} = \tau_1 \tau_2 +\tau_1 \tau_3+ \tau_2\tau_3$, where $\tau_1$, $\tau_2$ and $\tau_3$ are the eigenvalues of $\vec \tau$. 

To represent $\tr ( \ini {\vec B} \vec C)$ in terms of the invariants of $\vec \tau$ and $\vec C$, 
we multiply each side of Eq.~\eqref{eqn:tau} on the right with $\vec C$ and take the trace to get
\be
\label{eqn:trBetau}
  \mu \tr ( \ini {\vec B} \vec C)= \tr( \vec \tau \vec C ) + \ini p \tr \vec C     ,
\en
which we use to write the free-energy density Eq.\eqref{eqn:NeoHookPsi} as
\be
  \Psi = \frac{1}{2} \left( \ini p I_1 + J_1 -3\mu \right)
  \label{eqn:PsiInTau}
\en
with $\ini p$ given by Eq.~\eqref{eqn:inip}${}_1$. Note that in the absence of residual stress $\vec \tau =0$,  $\ini p = \mu$ by~\eqref{eqn:inipCubic}, $J_1 = I_1$, and then $\Psi$ reduces to the classical neo-Hookean model, as expected.  Equation \eqref{eqn:PsiInTau} represents the general extension of the neo-Hookean strain energy function to a residually stressed material, resulting in a function of only five of the nine independent invariants of  ${\vec C}$ and ${\vec \tau}$. 
%We see that even for a Neo-Hookean material the derivative $\Psi_{I_1}$ is an elaborate function of $\vec \tau$'s invariants. 

Substituting~\eqref{eqn:PsiInTau} in the Cauchy stress Eq.~\eqref{eqn:CauchyStressGeneral} we arrive at the constitutive relation
\be
\label{eqn:SigmaInTau}
\vec \sigma = \ini p \vec B + \vec F \vec \tau \vec F^T - p \vec I.
\en
Taking $\bb$ as the reference configuration, $\ini \bb$ as the current configuration and leaving $\tilde \bb$ as the stress-free configuration, see Figure~\ref{fig:Configurations}, the initial stress becomes $\vec \sigma$, the Cauchy stress becomes $\vec \tau$, and Eq.~\eqref{eqn:SigmaInTau} becomes 
%this Cauchy stress was deduced from a virtual stress-free configuration, we know that is satisfies ISS~\eqref{cond:ISS}, see Appendix~\ref{app:VirtualConfig},
 %implying that
\be
\label{eqn:TauInSigma}
\vec \tau = \ini p_\tau \vec C^{-1} + \vec F^{-1} \vec \sigma \vec F^{-T} - p_\tau \vec I,
\en
where $\ini p_\tau$ is given by replacing $\vec \tau$ for $\vec \sigma$ in Eq.~\eqref{eqn:inip}${}_1$, and $p_\tau$ is an undetermined scalar.

%These two identities can also be achieved by calculating $\vec \sigma$ directly from~\eqref{eqn:NeoHookPsi} and then taking the determinant on either side.
%An even simpler model results from assuming that the residual stress was due to the deformation of a compressible material with $\Psi = \mu/2 \tr \tilde{\vec C}$, the result would be
%\be
  %\Psi= \Psi( J_1) = \frac{1}{2} \left(J_1  -3\mu \right),
  %\label{eqn:PsiInTau2}
%\en 
%with the Cauchy stress given by
%\bal
%\label{eqn:SigmaCompress}
%&\vec \sigma = \vec F \vec \tau \vec F^T.
%\eal
%The constitutive choices~\eqref{eqn:SigmaInTau} and~\eqref{eqn:SigmaCompress} satisfy the three requirements.             
%\subsubsection*{Neo-Hookean ISS}

 Substituting~\eqref{eqn:SigmaInTau} in~\eqref{eqn:TauInSigma} we find the connection
\be
(\ini p_\tau - p) \vec C^{-1} = ( p_\tau - \ini p) \vec I.
\en  
 As this equation must hold for every $\vec C$ we conclude that 
\be
\label{eqns:ps}
\ini p_\tau = p \quad \text{ and } \quad p_\tau = \ini p.
\en
Note that, as is expected of a neo-Hookean material, the above equations do not determine $p$ in terms of $\vec \tau$ or $\vec F$.
 %we know this as the constitutive choice~\eqref{eqn:SigmaInTau} was deduced from a neo-Hookean material. 
 %Example~\ref{subsec:example} in Section~\ref{sec:independentJ_3&J4} discusses a constitutive choice where $p$ is determined in terms of $\vec \tau$ and $\vec F$. 
%For the constitutive choice \eqref{eqn:SigmaInTau} to be physically viable, the above Eqs. must hold and $p$ should not be determined from $\vec \tau$ or $\vec F$, which we will demonstrate below. 
%Here we will show that $\ini p_\tau = p$ does not  $p$

%Rearranging Eq.~\eqref{eqn:SigmaInTau} we get 
%\be
%\vec \sigma + p \vec I =  \vec F (\vec \tau + \ini p \vec I  ) \vec F^T.
%\en
%Taking the determinant on either side and using $\det F = 1$ we reach 
%\be
%\det (\vec \sigma + p \vec I ) =  \det (\vec \tau + \ini p \vec I  ),
%\en
%where the right hand-side 

%Setting $\vec F = \vec I$ and $\vec \tau = \vec \sigma$ in Eq.~\eqref{eqn:SigmaInTau} we find that $p = \ini p$  

 %which could also be concluded from the ISS Eqs.~\eqref{eqns:ISSJ1}${}_3$ and~\eqref{eqns:ISSJ1}${}_4$. 
 
\subsection{ Plane Strain}
The free energy density~\eqref{eqn:PsiInTau} is simplified when the residual strain has only planar components. Accordingly, let us assume $\ini B_{13}=\ini B_{23} =0$ and $\ini B_{33} =1$, which substituted in Eq.~\eqref{eqn:tau} results in $\tau_{13}=\tau_{23} =0$ and $\tau_{33} = \mu  - \ini p$. We let ${\vec B}_P$, ${\vec C}_P$, $ {\ini {\vec B}_P}$, $\vec I_P$, $\vec \sigma_P$ and $ {\vec \tau}_P$ be ${\vec B}$, ${\vec C}$, $ {\ini {\vec B}}$, $\vec I$, $\vec \sigma$ and $ {\vec \tau}$ restricted to the $(x_1,x_2)$ plane respectively. From equation~\eqref{eqn:tau} we get
\be
\label{eqn:tau2D}
  \mu  \ini{\vec B}_P= {\vec \tau}_P +  \ini p {\vec I}_P.
\en  
 %which we  solve for $\tr \hat B$ in the case that $\tilde \Psi'(\tr \hat B) = \mu  = \mu +\kappa + \kappa \tr \hat B$, to obtain
%\bga
%\tilde \Psi' = \frac{\kappa}{2}+\frac{\mu}{2}  +\frac{1}{2}\sqrt{(k+\mu )^2+2 k (2 \ini p+ \tr \hat \tau )}.
%\ega
%Returning to Eq.~\eqref{eqn:tau2D}, 
To obtain $\ini p$ in terms of $\vec \tau_P$ we take the determinant of each side of Eq.~\eqref{eqn:tau2D} giving
\bga
 \det \left ( \mu \ini {\vec B}_P \right ) = \det (  {\vec \tau}_P +\ini p  {\vec I}_P ) \implies  \mu^2  = \ini p^2 + \ini p \tr  {\vec \tau}_P  + \det  {\vec \tau}_P
\ega 
where we used $\det  (\mu \ini {\vec B} ) =\mu^2$. We solve the above for $\ini p$ to get
\bga
\label{eqn:p2D}
\ini p_{\pm} = -\frac{\tr  {\vec \tau}}{2} \pm \frac{1}{2}\sqrt{-4 \det  {\vec \tau} +( \tr  {\vec \tau})^2 + 4 \mu^2 },
\ega
where $-4 \det  {\vec \tau} +( \tr  {\vec \tau})^2 + 4 \mu^2 = (\tau_1-\tau_2)^2 +4 \mu^2$ in terms of the eigenvalues of $\vec \tau$. The solution $\ini p=\ini p_{-}$ should be discarded due to the following: $\ini p_{-} = - \mu$ when $\vec \tau_P =\vec 0$, and Eq.~\eqref{eqn:tau2D} shows that $\ini p$ should be positive when $\vec \tau_P =\vec 0$, so for $\vec \tau_P =\vec 0$ the only viable solution is $\ini p = \ini p_{+}$. Further, as we expect $\ini p$ to be continuous in $\vec \tau_P$ for every $\vec \tau_P \in \mathbb R^2$, and $\ini p_{-} \not = \ini p_{+}$ for every $\vec \tau_P$, we should discard the solution $\ini p_{-}$.     

For simplicity we assume $C_{32}= C_{31} =0$ and then Eq.~\eqref{eqn:NeoHookPsi} reads 
\be
%\Psi = \frac{\mu}{2}\left(  \tr \tilde{\vec B} -3 \right) -\frac{\mu}{2} = \frac{\mu}{2} \left ( \tr(  \ini {\vec B} {\vec C}  ) -2 \right) + \frac{\mu}{2}(C_{33} -1),
\Psi = \frac{\mu}{2} \left ( \tr(  \ini {\vec B_P} {\vec C_P}  ) -2 \right) + \frac{\mu}{2}(C_{33} -1),
\en
 and substituting $\ini {\vec B_P}$ from Eq.~\eqref{eqn:tau2D} we get 
%{\vec \tau}_P/\mu + \ini p {\vec I}_P/\mu
%We take the trace of both sides and rearrange to obtain $ \tr \hat{\vec  B_P}  =  {\vec \tau}_P/ \mu + 2\ini p/\mu $.
\be
   \Psi =    \frac{1}{2} \tr( \vec \tau_P \vec C_P)+ \frac{1}{2} \ini p \tr \vec C_P   - \mu + \frac{\mu}{2}(C_{33} -1),
  \label{eqn:Psi2D}
\en
where $\ini p = \ini p_{+}$ given by Eq.~\eqref{eqn:p2D}.
The Cauchy stress Eq.~\eqref{eqn:CauchyStressGeneral} becomes,
\bga
\label{eqn:sigma2D}
\vec \sigma_P = \ini p  \vec B_P - p \vec I_P + \vec F_P \vec \tau_P \vec F^T_P, 
\ega  
and $\sigma_{33} = \ini p - p +\tau_{33} = \mu - p$, where we have used that $\tau_{33}= \mu - \ini p$.
%We can deduce the only out of plane component $\sigma_{33}$ separately by using $\tau_{33}= \mu - \ini p$, so that $\sigma_{33} = \mu B_{33} \ini B_{33} - p =  B_{33}(\tau_{33} +\ini p) - p $. 
For $\vec F = \vec I$ we have $\vec \sigma = \vec \tau$ and $ \ini p =p$.

%\subsection{Uniaxial strain example}
%\label{sec:Uniaxial}

%A simple example to examine the neo-Hookean model is in an infinite space with $F_{11} = F_{33} = \lambda^{1/2}$, $F_{22} = \lambda^{-1}$ and $\tau_{ij} = \tau_i \delta_{ij}$.

%If $\tau_1=\tau_2=\tau_3$ then $\ini p = -\tau_1$. 
 
%If $\tau_1=\tau_2$ then $T_2 = -27$.  

%$\vec \tau = \mu \vec B - \ini p \vec I \implies \tau _1 = \mu \lambda_1^2 -p, \quad \tau_2 = \mu \lambda_2^2 -p, \quad \tau _3 = \frac{\mu}{\lambda_1^2 \lambda_2^2} -p.$ So naturally no restrictionso on the $\tau_j$'s

\section{The Initial Stress Symmetry Equations}
\label{sec:ISS}
In this Section we explained how to express the ISS condition~\eqref{cond:ISS} as 
a set of scalar equations that relates the free energy density $\Psi$; $p$ and $p_\tau$ from Eq.~\eqref{cond:ISS}; and the invariants of $\vec \tau$ and $\vec C$. Let us first consider a simple example, namely assuming $\Psi = I_1 J_1/2$. One can check that this strain energy satisfies the compatibilty equations~\eqref{eqn:StressCompatibility} with $\ini p = \tr \vec \tau$, but does it satisfy the ISS~\eqref{cond:ISS}? We can use the stress in the form~\eqref{eqn:CauchyStressGeneral} to write
\be
\label{eqn:ISSexample1}
\vec \sigma = \hat{\vec \varsigma}(\vec F,\vec \tau, p) =  \tr (\vec \tau \vec C ) \vec B -p \vec I + \tr(\vec C) \vec F \vec \tau \vec F^T,
\en
and then from ISS we have that
\be
\label{eqn:ISSexample2}
\vec \tau = \hat{\vec \varsigma}(\vec F^{-1},\vec \sigma, p_\tau) =  \tr ( \vec \sigma \vec B^{-1}) \vec C^{-1} -p_\tau \vec I + \tr\left( \vec B^{-1}\right) \vec F^{-1} \vec \sigma \vec F^{-T}.
\en
To check that both Eqs.~\eqref{eqn:ISSexample1} and~\eqref{eqn:ISSexample2} hold for every $\vec \tau$ and $\vec F$ we substitute $\vec \sigma$ into Eq.~\eqref{eqn:ISSexample2}, which after some rearranging becomes
\be
\label{eqn:ISSexample3}
(1-I_1 I_2 )\vec \tau =   ( 3 J_1 -2 p I_2 + I_1 I_{\tau1} ) \vec C^{-1}  + (I_2 J_1  -p_\tau) \vec I   ,
\en
where we have used that $\tr (\vec B^{-1}) = I_2$, which can be shown by applying the Cayley-Hamilton theorem to $\vec B$ with $I_3 =1$ (due to incompressibility). In order to satisfy Eq.~\eqref{eqn:ISSexample3} for every $\vec F$ and $\vec \tau$, the coefficients of $\vec \tau$, $\vec C^{-1}$ and $\vec I$ must all be identically zero (this is shown rigorously in Appendix~\ref{app:ReduceISS}). For the coefficient of $\vec \tau$ to be zero, the identity $I_1 I_2 =1$ must hold for every $\vec F$, which is obviously impossible: so we conclude that $\Psi = I_1 J_1/2$ does not correctly furnish the Cauchy stress for every reference configuration.  

In Appendix~\ref{app:ReduceISS} we reduce ISS to nine scalar equations, following a procedure similar to the one above. They are compactly written as
\bal
\label{eqn:ISS1}
& \vec b
+
\vec P^{\{1\}}\Psi_{J_1}^\sigma + \vec P^{\{2\}}\Psi_{J_2}^\sigma
+  \Psi_{J_n} \left ( \vec P_{\{n\}}^{\{3\} }  \Psi_{J_3}^\sigma +   \vec P_{\{n\}}^{\{4\} }  \Psi_{J_4}^\sigma  \right )=0,
\\
\label{eqn:ISS2}
& \Psi_{J_n} \left (  \vec Q_{\{n\}}^{\{1\} }  \Psi_{J_1}^\sigma +\vec Q_{\{n\}}^{\{2\} }  \Psi_{J_2}^\sigma 
%\left( \Psi_{J_3} \vec Q_{\{3\}}^{\{1\} } +\Psi_{J_4} \vec Q_{\{4\}}^{\{1\} } \right) \Psi_{J_1}^\sigma +\left( \Psi_{J_3} \vec Q_{\{3\}}^{\{2\} } +\Psi_{J_4} \vec Q_{\{4\}}^{\{2\} } \right) \Psi_{J_2}^\sigma 
+     \vec Q_{\{n\}}^{\{3\} }  \Psi_{J_3}^\sigma +  \vec Q_{\{n\}}^{\{4\} }  \Psi_{J_4}^\sigma \right) =0,
\eal
where $n$ sums over $0,1,2,3,4$, $\Psi_{J_0} \coloneqq 1$,  
\bal
& \Psi_{I_k} = \Psi_{I_k} (\vec F, \vec \tau), \quad \Psi_{J_m} = \Psi_{J_m} (\vec F, \vec \tau),
\\
& \Psi_{I_k}^\sigma \coloneqq \Psi_{I_k} (\vec F^{-1}, \vec \sigma), \; \Psi_{J_m}^\sigma \coloneqq \Psi_{J_m}(\vec F^{-1}, \vec \sigma), 
\eal
for $k  \in \{1, 2\}$ and $m \in \{ 1,2,3, 4\}$ with $\vec \sigma$ given by~\eqref{eqn:CauchyInvariants}, \begin{gather}
 \vec b = 
\Scale[0.8]{
\begin{bmatrix}
2 \Psi_{I_2}^\sigma \\
-2 \Psi_{I_1}^\sigma \\
p_\tau \\
0 \\
0 \\
1
\end{bmatrix}
},
\; 
\vec P^{\{1\}}=
4
\Scale[0.8]{
\begin{bmatrix}
\Psi_{I_2}
  \\
p/2  \\
-\Psi_{I_1} \\
-2 \Psi_{J_2} \\
2 I_2 \Psi_{J_2}  \\
   - \Psi_{J_1} - 2 I_1 \Psi_{J_2} \\
\end{bmatrix}
}
,
 \; 
\vec P^{\{2\}}=
4
\Scale[0.8]{
\begin{bmatrix}
  p \\
 I_2 p - 2 \Psi_{I_1} -2 I_1 \Psi_{I_2} \\
 2 \Psi_{I_2} \\
 0 \\
  -2 \Psi_{J_1}  \\
 -4 \Psi_{J_2}
\end{bmatrix}
},
\\
 \vec Q^{\{1\}}_{\{1\}}=\vec Q^{\{1\}}_{\{2\}}=\vec Q^{\{2\}}_{\{1\}}=\vec Q^{\{2\}}_{\{2\}} = \vec 0,
\end{gather}
 while all over matrices are given in Appendix~\ref{app:ISSmatrices}.  Eqs~\eqref{eqn:ISS1} and~\eqref{eqn:ISS2} represent 6 and 3 scalar equations, respectively, with the unknowns being the functions $\Psi$, $p$ and $p_\tau$. 

We now look at special cases of $\Psi$ which simplify the ISS equations and lead to more practical representations for free energy density.

\subsection{Free energy independent of $J_3$ and $J_4$}
 \label{sec:independentJ_3&J4}
 When $\Psi$ is independent of $J_3$ and $J_4$, Eq.~\eqref{eqn:ISS2} is identically zero and only the first three terms of Eq.~\eqref{eqn:ISS1} are non-zero. The fourth scalar equation of~\eqref{eqn:ISS1} becomes $-8 \Psi_{J_2}\Psi_{J_1}^\sigma =0$ which is true, for every $\vec F$ and $\vec \tau$, only when $\Psi$ does not depend on $J_2$ or does not depend on $J_1$. We will investigate the first case in more detail below. 

\subsubsection*{Assuming $\Psi$ independent of $J_2$}
If $\Psi$ does not depend on $J_2$, $J_3$ and $J_4$ then Eq.~\eqref{eqn:ISS1} reduces to
%\be
 %4 \Psi_{J_1} \Psi_{J_1}^{\vec \sigma} =1, 
 %\;\; 
 %\Psi_{I_2}^{\vec \sigma} + 2 \Psi_{I_2} \Psi_{J_1}^{\vec \sigma} =0, 
 %\;\;
 %- \Psi_{I_1}^{\vec \sigma} +  p \Psi_{J_1}^{\vec \sigma} =0, 
%\;\;
 %4 \Psi_{I_1} \Psi_{J_1}^{\vec \sigma} = p_{\vec \tau},
 %\label{eqns:ISSJ1}
%\en
\be
\label{eqns:ISSJ1}
 4 \Psi_{J_1} \Psi_{J_1}^{\vec \sigma} =1, 
 \;\; 
 \frac{\Psi_{I_2}^{\vec \sigma}}{\sqrt{\Psi_{J_1}^{\vec \sigma}}} + \frac{\Psi_{I_2}}{\sqrt{\Psi_{J_1}}}  =0, 
 \;\;
    p \Psi_{J_1}^{\vec \sigma} = \Psi_{I_1}^{\vec \sigma}, 
\;\;
  p_{\vec \tau} \Psi_{J_1} =\Psi_{I_1} , 
\en
where we have used Eq.~\eqref{eqns:ISSJ1}${}_1$ to derive the other three equations.
 We can check if these equations are consistent with the compatibility Eqs.~\eqref{nc} by letting $\vec F =\vec I$, $\vec \sigma = \vec \tau$ and $p_\tau = p =\ini p$. This results in $\Psi_{J_1} =\pm 1/2$, $2 \Psi_{I_1} = \pm \ini p$ and $\Psi_{I_2}(1 \pm 1) =0$, which only satisfies Eqs.~\eqref{nc} if $\Psi_{J_1} = 1/2$, $2\Psi_{I_1} =  \ini p$ and $\Psi_{I_2} =0$ for $\vec F =\vec I$. 

%For simplicity we choose $\Psi_{J_1} = 1/2$ implying that $\Psi_{J_1}^\sigma = 1/2$ for all $\vec F$. With this choice Eq.~\eqref{eqns:ISSJ1}${}_2$ becomes 
%\[
%\Psi_{I_2}(I_2,I_1,\vec \sigma) + \Psi_{I_2}(I_1,I_2,\vec \tau)=0, 
%\] 
%where we have used that $I_1$ is swapped for $I_2$ when $\vec F^{-1}$ is substituted for $\vec F$. To satisfy the above Eq. it is simpler to drop the dependance of $\Psi_{I_2}$ on the stress and choose a strain energy in the form
%\be
%\label{eqn:PsiJ1}
%\Psi = \frac{1}{2}f(I_1, \vec \tau) + \frac{1}{2}g(I_1 - I_2) + \frac{J_1}{2},
%\en 
%where $g>0$ and $g'(I_2 - I_1) = - g'(I_1 - I_2)$, then Eqs.~\eqref{eqns:ISSJ1} and~\eqref{eqn:CauchyInvariants} give
%\bal
%\label{eqn:PsiJ1_ISS}
%&p = \frac{\partial f}{\partial I_1}(I_2,\vec \sigma) -g'(I_1 - I_2), \;\; p_\tau = \frac{\partial f}{\partial I_1}(I_1,\vec \tau) + g'(I_1 - I_2) \;\; \text{and}
%\\
%\label{eqn:PsiJ1_sigma}
%&\vec \sigma = \frac{\partial f}{\partial I_1}(I_1,\vec \tau) \vec B - g'(I_1 -I_2)((I_1 -1)\vec B  -\vec B^2)  -p \vec I + \vec F \vec \tau \vec F^T.
%\eal
%which must hold for every $\vec \tau$ and $\vec F$. Note that if $f$ did not depend on $\vec \tau$ then both $p$ and $p_\tau$ would be completely determined from the deformation, which would pose a challenge in many applications.
Let us apply Eqs.~\eqref{eqns:ISSJ1} to the following example 
\be
\label{eqn:MerodioPsi}
\Psi = \frac{1}{2} \mu (I_1 -3) + \frac{1}{2}(J_1 - I_{\tau_1}),
\en
 used by \citeauthor{merodio2013influence}(\citeyear{merodio2013influence}). In this case, Eqs.~\eqref{eqns:ISSJ1}${}_1$ and~\eqref{eqns:ISSJ1}${}_2$ are satisfied while Eqs.~\eqref{eqns:ISSJ1}${}_3$ and~\eqref{eqns:ISSJ1}${}_4$ become $p = p_\tau = \mu$. However, restricting $p$ to be constant, which is normally determined through boundary conditions, would result in physically unexpected behaviour. This is best seen with an example.

%\subsubsection{ example}
\begin{example}
\label{subsec:example}
For the free energy density expressed in Eq.~\eqref{eqn:MerodioPsi} the Cauchy stress from Eq.\eqref{eqn:CauchyInvariants} becomes
\be
\label{eqn:CauchyJose}
\vec \sigma = \mu \vec F^T \vec F - \vec I p + \vec F^T \vec \tau \vec F.
\en
Let us consider a cube in the reference configuration, subject to the residual stress $\tau_{ij} = \tau_i \delta_{ij}$, aligned with the axes of the cube.  For the current configuration we impose two clamped conditions  that fix $F_{11} = \lambda_1$ and $F_{33} = \lambda_3$, then $F_{22} = \lambda_2 = (\lambda_1 \lambda_3)^{-1}$ due to incompressibility, with all the other components of $\vec F$ being zero. With this imposed deformation the Cauchy stress is diagonal with components $\sigma_{ij} = \delta_{ij} \sigma_i$.  We can also prescribe the stress $\sigma_{2}$, as this will not alter $\lambda_1$ and $\lambda_3$ because they are kept fixed. We choose $\sigma_{2} = \tau_{2}$, to ensure compatibility with the reference configuration when $\lambda_1=\lambda_3 =1$, which results in 
\[
\tau_2 = \mu \lambda_2^2 -p  +\lambda_2^2 \tau_2.
\]
%which can be used to determine $p$.
%The stretch $\lambda_2$ is given in terms of $\lambda_1$ and $\lambda_3$, so we can view $p$ as being determined by $\lambda_1$, $\lambda_3$ and $\tau_2$. 
Clearly it is not possible for $p$ to satisfy the above and be fixed at $p=\mu$ so that ISS is satisfied. This means that the free energy density Eq.~\eqref{eqn:MerodioPsi} either results in non-physically behaviour, or it does not satisfy ISS; which implies that the constitutive relation~\eqref{eqn:CauchyJose} does not hold for every reference configuration.
\end{example}

The choice of $\Psi$ should satisfy Eqs.~\eqref{eqns:ISSJ1} without restricting the deformation $\vec F$, initial stress $\vec \tau$ or $p$, otherwise the material will likely exhibit non-physical behaviour. Below we deduce a a free energy density $\Psi$ for a compressible model, inspired by the neo-Hookean example in Section~\ref{sec:SimpleEnergy}, that satisfies ISS without any unphysical restrictions.  

For a compressible material we substitute $p$ with $- 2 I_3 \Psi_{I_3}$ and $p_\tau$ with $- 2 I_3^{-1} \Psi_{I_3}^{\vec \sigma}$ in the ISS Eqs.~\eqref{eqns:ISSJ1}. We will find under what conditions does the following free energy density
\be
\Psi = \frac{1}{2}g (\vec \tau, I_3) I_1 I_3^{-1/3} + \frac{1}{2}f(\vec \tau,I_3) + \frac{J_1}{2} I_3^{-1/3} 
\label{eqn:PsiCompressibleNeoHook}
\en 
satisfy ISS~\eqref{eqns:ISSJ1} and stress compatibility~\eqref{nc}, where $f$ and $g$ are arbitrary scalar functions.

The Cauchy stress~\eqref{eqn:CauchyInvariants} becomes
\be
\label{eqn:neoHookCompressCauchy}
\vec \sigma = g(\vec \tau, I_3) I_3^{-1/3} \vec B + 2 I_3 \Psi_{I_3} \vec I + I_3^{-1/3} \vec F^T \vec \tau \vec F, 
\en
with 
\be
 \label{eqn:neoHookPsiI3}
 \Psi_{I_3} =  \frac{1}{2}f_{I_3}(\vec \tau, I_3)+ \frac{1}{2}g_{I_3}(\vec \tau, I_3) I_1 I_3^{-1/3} - \frac{1}{6} I_3^{-4/3}\left [g (\vec \tau, I_3) I_1 + J_1 \right ]. 
\en
The ISS Eq.~\eqref{eqns:ISSJ1}${}_1$ is satisfied as $\Psi_{J_1} \Psi_{J_1}^{\vec \sigma} = (1/2)I_3^{-1/3} (1/2)I_3^{1/3}  =  1/4$ and stress compatibility~\eqref{nc}${}_2$ is satisfied as $2\Psi_{J_1} = 1$ when $\vec F = \vec I$. By substituting $p$ for $-2 I_3 \Psi_{I_3}$, the remaining compatibility Eq.~\eqref{nc}${}_1$ becomes 
\bga
\label{eqn:neoHookCompat_1}
g(\vec \tau, I_3) = - 2 \Psi_{I_3} \quad \text{evaluated at} \quad \vec F = \vec I.
\ega
Eq.~\eqref{eqn:neoHookCompat_1} is satisfied if we set $f_{I_3}(\vec \tau, 1)  = \tr \vec \tau/3- 3 g_{I_3}(\vec \tau, 1)$, then Eq.~\eqref{eqn:neoHookCompat_1} and Eq.~\eqref{eqn:neoHookCompressCauchy} together imply that $\vec \sigma = \vec \tau$ when $\vec F = \vec I$.

The ISS Eqs.~\eqref{eqns:ISSJ1}${}_3$ and~\eqref{eqns:ISSJ1}${}_4$ now read respectively 
\be
 -2 I_3 \Psi_{I_3} =  g(\vec \sigma, I_3^{-1}) \;\;\text{ and }\;\;  -2 I_3^{-1} \Psi_{I_3}^{\vec \sigma} =  g(\vec \tau, I_3), 
 %\quad \text{for every} \;\; \vec \tau \; \;\text{and} \;\; \vec F,
\label{eqns:ISSJ1_34}
\en
for every $\vec F$ and $\vec \tau$, 
where $f_{I_3}$ and $g_{I_3}$ are respectively the partial derivatives of $f$ and $g$ with respect to $I_3$. Note that the above two equations are equivalent as Eq.~\eqref{eqns:ISSJ1_34}${}_1$ becomes Eq.~\eqref{eqns:ISSJ1_34}${}_2$ when we substitute $\vec F$ for $\vec F^{-1}$ and $\vec \tau$ for $\vec \sigma$. For this reason it is sufficient to satisfy Eq.~\eqref{eqns:ISSJ1_34}${}_1$. If $\vec F =\vec I$, Eq.~\eqref{eqns:ISSJ1_34}${}_1$ is satisfied as it reduces to Eq.~\eqref{eqn:neoHookCompat_1}. The rest of this section will focus on showing that Eq.~\eqref{eqns:ISSJ1_34}${}_1$ is satisfied for every $\vec F$ and $\vec \tau$. 

By rearranging Eq.~\eqref{eqn:neoHookCompressCauchy} and taking the determinant on either side we get
%\[
%\det \left(\vec \sigma - 2 I_3 \Psi_{I_3} \vec I \right ) = \det \left (   2\Psi_{J_1} \vec F^T \vec \tau \vec F + 2 \Psi_{I_1} \vec B  \right )  =
  %(2\Psi_{J_1})^3 \det \vec B \det (\vec \tau + \Psi_{I_1}/ \Psi_{J_1} \vec I   )
%\]
\[
\det \left(\vec \sigma - 2 I_3 \Psi_{I_3} \vec I \right ) = \det \left (   I_3^{-1/3}  \vec F^T \vec \tau \vec F + g (\vec \tau, I_3) I_3^{-1/3}  \vec B  \right )  =
  I_3^{-1} \det \vec B \det (\vec \tau + g(\vec \tau, I_3) \vec I   ),
\]
and by using incompressibility $\det \vec B =1$ 
%and Eq.~\eqref{eqns:ISSJ1}${}_1$ 
we find that 
\be
\label{eqn:DetOfCompressible}
I_3^{1/2} \det (\vec \sigma - 2 I_3 \Psi_{I_3} \vec I) =   I_3^{-1/2} \det (\vec \tau + g(\vec \tau, I_3) \vec I).
\en
If we choose $g(\vec \tau,I_3)$ such that
%to be one of the possibly three real solutions to 
\be
 I_3^{-1/2} \det (\vec \tau +  g(\vec \tau,I_3)   \vec I) = k,
 \label{eqn:Choose_g}
\en
%then $g(\tau, I_3)$ will be one of the three solutions~\eqref{eqn:inip} to $\ini p$ but with $\mu^3$ replaced by $2^{3/2} k I_3^{1/2} $.
%If we now restrict $\Psi$ so that 
%\be
%\label{eqn:ChooseConstant}
%(\Psi_{J_1})^{3/2} \det (\vec \tau + \Psi_{I_1}/ \Psi_{J_1} \vec I) = k
%\en
where $k$ is a constant, then the right handside of Eq.~\eqref{eqn:DetOfCompressible} is also equal to $k$, thus $\vec \sigma$ must satisfy
%the left-hand side of~\eqref{eqn:DetOfCompressible} will also equal $k$. So $\vec \sigma$ must satisfy
\be
I_3^{1/2} \det (\vec \sigma - 2 I_3 \Psi_{I_3} \vec I) =k.
\label{eqn:ChoosePsi3}
\en
By swapping $\vec \tau$ for $\vec \sigma$ and $\vec F$ for $\vec F^{-1}$ Eq.~\eqref{eqn:Choose_g} becomes
\be
I_3^{1/2} \det (\vec \sigma +  g(\vec \sigma,I_3^{-1})   \vec I) = k.
\label{eqn:Choose_g2}
\en
For $g(\vec \tau,I_3)$ to satisfy Eq.~\eqref{eqn:Choose_g} there are three possible solutions given by replacing $\mu^3$ with $k I_3^{1/2}$ and $g(\vec \tau,I_3)$ with $\ini p$ in~\eqref{eqn:inip}${}_1$,\eqref{eqn:inip}${}_2$ and \eqref{eqn:inip}${}_3$, which we respectively denote by $g_1(\vec \tau,I_3)$, $g_2(\vec \tau,I_3)$ and $g_3(\vec \tau,I_3)$. We highlighted in Section~\ref{sec:SimpleEnergy} that only $g(\vec \tau,I_3)=g_1(\vec \tau,I_3)$ is a physically viable solution. So if we choose $g(\vec \tau,I_3)= g_1(\vec \tau,I_3)$, then for Eq.~\eqref{eqns:ISSJ1_34}${}_1$ to be satisfied we need
\be
\label{eqn:PsiI3_gj}
 g_1(\vec \sigma,I_3^{-1}) = - 2 I_3 \Psi_{I_3} \quad  \text{for every} \; \; \vec F \; \; \text{and} \; \; \vec \tau.
\en
 We will show that the above is a consequence of Eqs.~\eqref{eqn:ChoosePsi3}, \eqref{eqn:Choose_g2} and stress compatibility~\eqref{eqn:neoHookCompat_1}. Due to Eq.~\eqref{eqn:neoHookCompat_1}, we know that initially for $\vec F= \vec I$ Eq.~\eqref{eqn:PsiI3_gj} is true. We can also conclude from Eqs.~\eqref{eqn:ChoosePsi3} and \eqref{eqn:Choose_g2} that $- 2 I_3 \Psi_{I_3}$ will equal either $g_1(\vec \sigma,I_3^{-1})$, $g_2(\vec \sigma,I_3^{-1})$ or $g_3(\vec \sigma,I_3^{-1})$ for every $\vec F$. Since $- 2 I_3 \Psi_{I_3}$ should be a continuous function of $\vec F$ and $\vec \tau$, it can not change from $g_1(\vec \sigma,I_3^{-1})$ to another solution $g_k(\vec \sigma,I_3^{-1})$ because it can be shown that $g_k(\vec \sigma,I_3^{-1}) \not = g_1(\vec \sigma,I_3^{-1})$ for every $\vec \sigma$ and $I_3$. 
 %To clarify, suppose that we parameterize a deformation $\gamma \to \vec F(\gamma)$ with $\vec F(0) = \vec I$. If we consider $\vec \tau$ fixed, then $\vec \sigma = \vec \sigma(\gamma)$ is also a function of $\gamma$. When $\gamma =1$ we have that $\vec \sigma(0) = \vec \tau$ and $g_1(\vec \sigma(0),1) = - 2 \Psi_{I_3}$. If for some $\gamma = \gamma_c$ we have that $g_1(\vec \sigma(\gamma_c), I_3^{-1})= g_k(\vec \sigma(\gamma_c),I_3^{-1})$, then for $\gamma > \gamma_c$ it is possible that $g_k(\vec \sigma(\gamma),I_3^{-1}) = - 2 I_3 \Psi_{I_3}$ and $g_1(\vec \sigma(\gamma),I_3^{-1}) \not = - 2 I_3 \Psi_{I_3}$. However, it can be shown that  $g_1(\vec \sigma,I_3^{-1})$ does not equal either $g_2(\vec \sigma,I_3^{-1})$ or $g_3(\vec \sigma,I_3^{-1})$ for any choice of $\vec \sigma$ and $I_3$. 
 For this reason if $g_1(\vec \tau,1) = - 2 \Psi_{I_3}$ for $\vec F =\vec I$, then $g_1(\vec \sigma ,I_3^{-1}) = - 2 I_3 \Psi_{I_3}$ for every $\vec F$.
 
 %On the other hand the functions $g_2(\vec \sigma,I_3^{-1})$ and $g_3(\vec \sigma,I_3^{-1})$ are equal only if $T_2/T_1^{3/2} = 1$, where in this section $T_2$ and $T_1$ are given by Eqs.~\eqref{eqn:ptauTs2} with $\mu^3$ replaced by $k I_3^{1/2}$ and $\vec \tau$ replaced by $\vec \sigma$. However, for $g_2(\vec \sigma,I_3^{-1})$ (or $g_3(\vec \sigma,I_3^{-1})$) to be real when evaluated at $\vec F = \vec I$ and at any given $\vec F = \vec F_0$, we need $T_2/T_1^{3/2} \leq 1 $ for $\vec F = \vec I$ and $\vec F= \vec F_0$. This means that if $g_2(\vec \sigma,I_3^{-1})$ (or $g_3(\vec \sigma,I_3^{-1})$) is real for $\vec F =\vec I$ and $\vec F = \vec F_0$, then it is always possible to choice a parameterization $\gamma \to \vec F(\gamma)$, where $\vec F(0) = \vec I$ and $\vec F(1) = \vec F_0$, such that $T_2/T_1^{3/2} < 1 $ for $0<\gamma <1$. Assuming 
 %$g_2(\vec \sigma,I_3^{-1})$ ($g_3(\vec \sigma,I_3^{-1})$) is   
%a viable real solution for $g(\vec \sigma,I_3^{-1})$, we conclude that if $g_2(\vec \tau,1) = - 2 \Psi_{I_3}$ ($g_3(\vec \tau,1) = - 2 \Psi_{I_3}$) for $\vec F =\vec I$, then $g_2(\vec \sigma ,I_3^{-1}) = - 2 I_3 \Psi_{I_3}$ ($g_3(\vec \sigma ,I_3^{-1}) = - 2 I_3 \Psi_{I_3}$) for every $\vec F$.

%For the incompressible neo-Hookean material, we saw in Section~\ref{sec:SimpleEnergy} that we could choose any of the three solutions for $\ini p$ and guarantee that $\ini p_\tau= p$. Here we want to show the analogous, that is, that we can choose any of the three solutions for $g(\vec \tau,I_3)$ and still guarantee that $g(\vec \sigma,I_3^{-1}) = - 2 I_3 \Psi_{I_3} $.

In conclusion, the model~\eqref{eqn:PsiCompressibleNeoHook} satisfies ISS as long as $g(\vec \tau,I_3)= g_1(\vec \tau,I_3)$, where $g_1(\vec \tau,I_3)$ is given by replacing $\mu^3$ with $k I_3^{1/2}$ in~\eqref{eqn:inip}${}_1$.

\section{Homogeneous stress gradient in a hollow cylinder}
\label{sec:HomogeneousStress}

It is now well acknowledged that residual stresses in living materials are vital to maintain ideal mechanical conditions. When an external load changes, growth and remodelling will alter the residual stress to best adapt to the new load.       
 %arise from a biological regulation process in order to adapt to external loads. 
%It is now understood that residual stress in biological tissue is vital to maintain ideal mechanical conditions. By adapting its residual stresses, the tissue can alter its internal stress and strain. 
An excellent example is how arteries remodel in response to the internal pressure.
The residual stress in arteries is thought to protect the arterial wall against strain concentration~\parencite{destrade2012uniform} or stress concentration \parencite{fung1991residual,taber1996theoretical}.
%Experimentally measurements have shown that when the blood flow increases, the arteries inner radius $a$ increases, but the thickness $b-a$ stays the same, but if only the pressure $P$ increases, then the arterial wall thickness increases $b-a$ while $a$ remains the same. 
%Allowing for residual stress lets the artery adapt to these changes while avoiding undesired stress or strain.
Here we will adopt the most accepted hypothesis, that the residual stress in the artery acts to minimize the stress gradient~\parencite{fung1991residual,cardamone2009origin}. The reasoning behind this hypothesis is that if a given tissue grows in response to stress, then homoeostasis is only possible if the tissue is under similar stress conditions throughout. So by minimizing the stress gradient we are selecting the most homogeneous stress possible. We will also consider a simplified artery with only one layer and no shear stress applied to the interior of the artery.
%\cite{polzer2013numerical}
%There is evidence that the material itself, governed in our case by $\mu$, does not change in response to either pressure or flow rate~cite ??. 

%Here we will not explore the process of growth itself. We simply assume that the body can grow into a state where the stress gradient is homogeneous. If an artery is to reach a certain ideal physiological state, such as a minimal stress gradient, this state needs to be feasible under the constitutive choice~\eqref{eqn:sigma2D}. Here we will show that...  To do so first we...

Our workflow is to first choose an optimal Cauchy stress field $\vec \sigma$, and then the to derive the residual stress from the ISS equation $\vec \tau = \hat{\vec \varsigma} (\vec F^{-1},\vec \sigma, p_\tau)$. So first we calculate the Cauchy stress with a near homogeneous circumferential and radial components
%\footnote{We note that the more common hypothesis is that the residual stress acts to homogenize the circumferential stress~\parencite{fung1991residual}.} 
in Section~\ref{sec:MinimalStressGrad}, then we find the residual stress that supports this optimal Cauchy stress in Section~\ref{sec:Inflation}. In Section~\ref{sec:MinimalStressGrad} we also show that the optimal Cauchy stress has a simple asymptotic formula. 

We finally compare the results with the corresponding ones obtained using the opening angle method~\parencite{chuong1986residual}. Pasquale: cute the rest.

\subsection{ Plain Strain Cylinder}
\label{sec:PlainStrain}
To describe the arterial wall we use cylindrical coordinates $(r,\theta, z)$ with the components of $\vec \sigma$ written in terms of unit basis vectors. We assume that the arterial wall retains its cylindrical symmetry when the internal pressure is removed and that there is no shear stress at the inner wall. These simplifications  
%are often made .. cite and 
imply that when a pressure is applied in the cylinder the resulting deformation is an inflation.
% What did I do in my previous more elaborate calculations on solving equilibrium Eqs. and boundary conditions for residually stressed cylindrically symmetric problems? I simply assumed that the residual stress was radially symmetric and that there was no shear stress, these two in turn guarantee that (for the neo-hook model) sigma is radially symmetric as well.

 %Due to radial symmetry and our assumptions we have that $\sigma_{zz}$, $\sigma_{zr}$ and  $\sigma_{z\theta}$ are all constant, $\vec \sigma$ is independent of $\theta$ and $z$, and the equilibrium Eqs. for the Cauchy stress reduces to
 Due to radial symmetry, the Cauchy stress $\vec \sigma$ is independent of $\theta$ and $z$, so in the absence of body forces the equilibrium equations reduce to
\bal
\label{eqn:equilibriumRadialCauchy1}
& \frac{\partial }{\partial r} \left ( r^2 \sigma_{\theta r} \right ) =0, \; \; \frac{\partial }{\partial r} \left ( r \sigma_{z r} \right ) =0,
\\
\label{eqn:equilibriumRadialCauchy2}
& \frac{\partial }{\partial r} \left ( r \sigma_{r r} \right )  -\sigma_{\theta \theta} =0,
\eal
 %\left.
%\begin{array}{c l}      
     %& \displaystyle  \frac{\partial}{\partial R} ( R^2  \tau^{\Theta R}  )  + R \frac{\partial \tau^{\Theta \Theta}}{\partial \Theta} = 0, \\
    %& \displaystyle \frac{\partial}{\partial R} (R \tau^{RR} ) + \frac{\partial \tau ^{R\Theta}}{\partial \Theta} - \tau^{\Theta \Theta}= 0,
%\end{array}\right \} 
%\be
%\label{eqn:sigmaCylindrical}
  %\displaystyle \frac{\partial}{\partial r} (r \sigma_{rr} )- \sigma_{\theta \theta}= 0
%\; \; \textrm{ for } a< r <b.
%\en
with the boundary conditions
\begin{align}
\label{eqn:sigmaCylindrical_boundarya}
    &  \; \sigma_{\theta r} =  0, \; \sigma_{z r} =  0 \; \textrm{ and } \; \sigma_{rr} =  -P \; \textrm{ for  } r = a, 
    \\ 
\label{eqn:sigmaCylindrical_boundaryb}
     &   \;  \sigma_{\theta r} =  0,  \;   \sigma_{z r} =  0 \; \textrm{ and } \; \sigma_{rr} =  0 \; \textrm{ for  } r = b,
\end{align}
where $a$ and $b$ are the inner and outer radius of the loaded artery, respectively. The equilibrium Eqs.~\eqref{eqn:equilibriumRadialCauchy1} together with the boundary conditions lead to $\sigma_{\theta r} =0$ and $\sigma_{z r} =0$ for all $r$. As neither $\sigma_{z\theta}$ nor $\sigma_{zz}$ appear in the equations of equilibrium, they are only restricted by the constitutive choice and boundary conditions on the cross-section of the artery. We use this degree of freedom to assume there is no axial torsion $\sigma_{z\theta} =0$.  
%while there may be a constant axial stress $\sigma_{zz}$. 
Note that for the constitutive choice Eq.~\eqref{eqn:CauchyInvariants} $\tau_{ZZ}$ can be chosen so that $\sigma_{zz}$ is constant, while for the simpler choice of plain strain~Eq.\eqref{eqn:sigma2D} (with $ \sigma_{zz}=\sigma_{33}$) $\sigma_{zz}$ is determined by $p$. Finally,
the remaining equation of equilibrium we need to enforce is Eq.~\eqref{eqn:equilibriumRadialCauchy2}. 

\subsection{Minimal stress gradient} 
\label{sec:MinimalStressGrad}
Here we develop a method to minimize the stress gradient fields $\sigma_{rr}'$ and $\sigma_{\theta \theta}'$, where the prime denotes differentiation in $r$. We make no assumptions about any reference configuration nor do we make any constitutive choice. We only make use of the assumptions from the section above which result in both $\sigma_{rr}$ and $\sigma_{\theta \theta}$ being independent of the coordinates $\theta$ and $z$; $\sigma_{\theta r}=  \sigma_{z r} =\sigma_{z \theta}= 0$; and for simplicity we will not consider $\sigma_{z z}$. %, as was done in cite.
 
%Below we will choose a $\sigma_{rr}$ that satisfies its boundary conditions~\eqref{eqn:sigmaCylindrical_boundarya} and~\eqref{eqn:sigmaCylindrical_boundaryb} and minimizes the stress gradient of $\sigma_{\theta \theta}$ and $\sigma_{rr}$ . 

Once $\sigma_{\theta \theta}$ and $\sigma_{rr}$ are determined, we use ISS~Eq.\eqref{eqn:ResidualISS} to write the residual stresses $\tau_{\Theta \Theta}$ and $\tau_{RR}$ in terms of $\sigma_{\theta \theta}$, $\sigma_{r r}$ and the deformation gradient in Section~\ref{sec:Inflation}.
 %At which point we make constitutive choices and solve the equilibrium Eq. and boundary conditions for $\vec \tau$.
 %rely on being able to choose any stress field $\sigma_{rr}$ that satisfies the boundary conditions~\eqref{eqns:CauchyBoundary} and equilibrium. As we saw in Section~\ref{sec:HomogeneousStress}, it is possible with our model of residual stress to choose $\sigma_{rr}'$ from which we obtain $\tau_{RR}'$ and then determine from the boundary condition~\eqref{eqn:DtauRBoundary} of $\tau_{RR}'$ one of, or a combination of, the parameters $a$, $b$, $A$ and $P$. It is likely that any model that allows to chose the residual stress field will also allow the same freedom of choice for $\sigma_{rr}$.

Using the equilibrium Eq.~\eqref{eqn:equilibriumRadialCauchy2} we write $\sigma_{\theta \theta}$ in terms of $\sigma_{rr}$ as 
\be
\label{eqns:ArteryCauchy}
\sigma_{\theta \theta} = \sigma_{rr} + r \sigma_{rr}' \implies \sigma_{\theta \theta}' = 2 \sigma_{rr}' + r \sigma_{rr}''. 
\en
As there are aortas and veins of many different sizes, for the sake of generality let us introduce the following dimensionless variables:
\be
\label{eqns:adimensional}
\varrho = \frac{r -a}{b-a} \quad \text{and} \quad \varsigma ( \rho) = \frac{\sigma_{rr}(r)}{P}, 
\en
which result in
\be
\label{eqns:adimensionalDvarrho}
 \sigma_{rr}' = \varsigma' \frac{P}{b-a}, \quad \sigma_{rr}'' = \varsigma'' \frac{P}{(b-a)^2},
\en
and
\be
\label{eqns:adimensionalDvarrho}
\sigma_{\theta\theta}' =  \frac{P}{b-a} \left (2 \varsigma' +  (\rho + \alpha)\varsigma''  \right ).
\en
Our aim is to minimize the stress gradient density  
\be
\label{normDsigma2}
 \frac{1}{b-a}\int_a^b \left [ \left (\sigma_{rr}' \right )^2 + (\sigma_{\theta \theta}')^2  \right ] dr =
\frac{P^2}{a^2}\alpha^2 \int_0^1 \left [  \left (\varsigma' \right )^2 + \left( 2 \varsigma' + \left ( \varrho + \alpha \right ) \varsigma''  \right )^2  \right ]  d\varrho 
\en
with the constraint
\be
\label{cons:varsigma}
 \varsigma(1)= 0 \; \text{ and } \; \varsigma(0) = -1 \implies \int_{0}^1 \varsigma' d \varrho =1,
\en
where we have introduced the quantity $\alpha = a(b-a)^{-1}$.

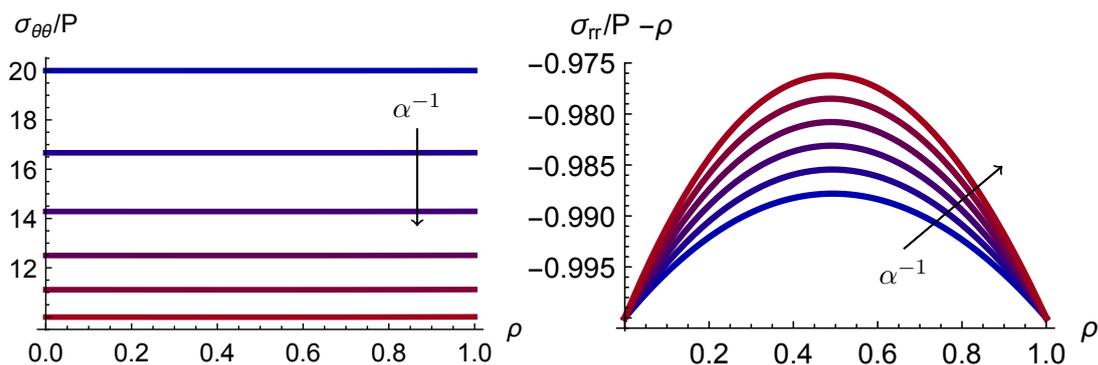
\begin{figure}[h!]  
\begin{tikzpicture} 
 	\draw (0,0) node {\includegraphics[width=0.45\textwidth]{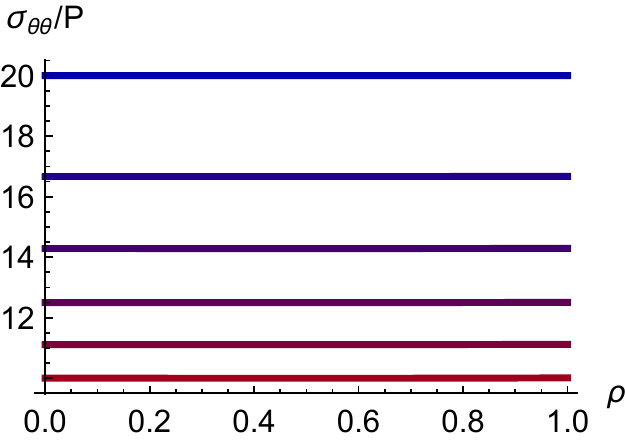}}; 
 	\draw (7.2,0) node {\includegraphics[width=0.5\textwidth]{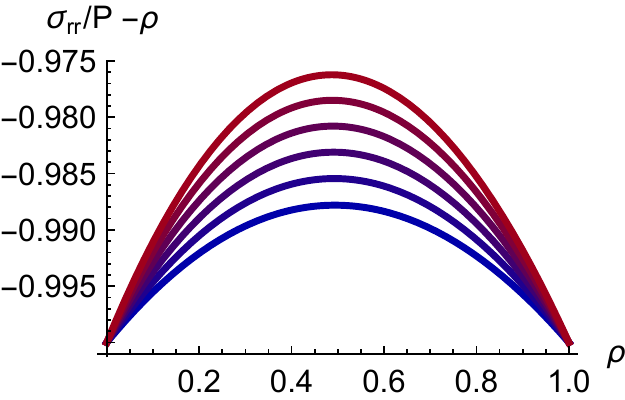}} ;
 	\draw (2.,0.8) node[above,thick]{$ \alpha^{-1}$};
	\draw [thick,->] (2.,.8) -- (2,-0.5);
 	\draw (8.4,-.8) node[below,thick]{$ \alpha^{-1}$};
	\draw [thick,->] (8.4,-.8)-- (9.7,0.3);
\end{tikzpicture}
\caption{The plots of the dimensionless Cauchy stress against the dimensionaless radius $\rho$, where the color shades from blue to red as $\alpha^{-1}$ (the wall thickness divided by the inner radius) goes from $0.05$ to $0.1$.}
\label{fig:SigmaHomo}
\end{figure}

Using the calculus of variations~\parencite{gregory1992constrained} we find that $\varsigma'$ 
%to minimize~\eqref{normDsigma2} with the above constraint it 
must satisfy the Euler-Lagrange equation
\be
\label{ode:EulerLagrange}
\frac{\partial f}{\partial \varsigma'} - \frac{d}{d \varrho }\frac{\partial f}{\partial \varsigma''} = \Lambda  \; \; \text{for} \; \varrho \in (0,1) \quad  \text{and} \quad \frac{\partial f}{\partial \varsigma''} = 0 \; \; \text{for} \; \varrho =0,1, 
\en
where 
\be
\label{eqns:f&alpha}
 f =   \left (\varsigma' \right )^2 + \left( 2 \varsigma' + \left ( \varrho + \alpha  \right ) \varsigma''  \right )^2, 
\en
and $\Lambda$ is a Lagrange multiplier due to the constraint~\eqref{cons:varsigma}. The solution to~\eqref{ode:EulerLagrange} is given by
\begin{multline}
\label{eqn:Dvarsigma}
 \varsigma' = \frac{\Lambda}{6} -\frac{\Lambda}{6} (\sqrt{13} -3) \frac{ \alpha^{\frac{\sqrt{13}}{2} + \frac{1}{2} } - (1+\alpha)^{\frac{\sqrt{13}}{2} + \frac{1}{2} } } {\alpha^{\sqrt{13}} - (1 +\alpha)^{\sqrt{13}}  }(\alpha+\varrho)^{\frac{\sqrt{13}}{2} + \frac{1}{2}}
 \\
 + \frac{2 \Lambda}{3 (\sqrt{13} -3)}  \frac{ \alpha^{\frac{\sqrt{13}}{2} - \frac{1}{2} } - (1+\alpha)^{\frac{\sqrt{13}}{2} -1 \frac{1}{2} } } {\alpha^{\sqrt{13}} - (1 +\alpha)^{\sqrt{13}}  } \left (\frac{\alpha (1+ \alpha )}{\alpha + \varrho} \right )^{\frac{\sqrt{13}}{2} + \frac{1}{2} }.
\end{multline}
We determine $\Lambda$ from the constraint~\eqref{cons:varsigma} to be
\begin{multline}
\frac{\Lambda}{6} =  \Big ( \int_0^1 1 - (\sqrt{13} -3) \frac{ \alpha^{\frac{\sqrt{13}}{2} + \frac{1}{2} } - (1+\alpha)^{\frac{\sqrt{13}}{2} + \frac{1}{2} } } {\alpha^{\sqrt{13}} - (1 +\alpha)^{\sqrt{13}}  }(\alpha+\varrho)^{\frac{\sqrt{13}}{2} + \frac{1}{2}}
 \\
 + \frac{4}{ \sqrt{13} -3}  \frac{ \alpha^{\frac{\sqrt{13}}{2} - \frac{1}{2} } - (1+\alpha)^{\frac{\sqrt{13}}{2} -1 \frac{1}{2} } } {\alpha^{\sqrt{13}} - (1 +\alpha)^{\sqrt{13}}  } \left (\frac{\alpha (1+ \alpha )}{\alpha + \varrho} \right )^{\frac{\sqrt{13}}{2} + \frac{1}{2} }d \varrho \Big)^{-1}.
\end{multline}
Realistic values for $\alpha = a (b-a)^{-1}$ can be obtained from in vivo measurements of the lumen radius (inner radius) $a$ and total artery thickness $b-a$. For the descending thoracic aorta of healthy men around 51 years old the mean lumen radius is $\SI{20}{\milli\meter}$~\parencite{stefanadis1995pressure} and the mean thickness is $\SI{1.4}{\milli\meter}$~\parencite{mensel2014mri}.
 %From~\cite{mensel2014mri} and~\cite{stefanadis1995pressure} we see that, for the descending thoracic, the mean thickness is 1.4mm and mean lumen of 20mm for men around 51 years old. 
 These estimates give $\alpha^{-1} \approx 0.07$ making it appropriate to expand $\varsigma'$ and $\Lambda$ as a power series in $\alpha^{-1}$, which gives
\bal
\varsigma'&= \frac{\Lambda}{2} \left ( 1 + (1-2\rho) \alpha^{-1} +\frac{1}{3} \left (9\rho^2 -6 \rho -1 \right )\alpha^{-2} \right )+ \bigO{\alpha^{-3}},
\\
\Lambda &= 2 + \frac{2}{3}\alpha^{-2} + \bigO{\alpha^{-3}},
\eal
or simply 
\be
\varsigma' = 1 +(1-2\rho) \alpha^{-1} +\rho(3 \rho -2)\alpha^{-2} + \bigO{\alpha^{-3}}.
\en
We can substitute $\varsigma'$ in Eqs.~\eqref{eqns:ArteryCauchy},~\eqref{eqns:adimensional} and~\eqref{eqns:adimensionalDvarrho} to obtain
\bal
\label{eqn:Cauchyr_homo}
\sigma_{rr} &= -P(1-\rho) (1 - \rho \alpha^{-1} +\rho^2 \alpha^{-2}) +\bigO{\alpha^{-3}},
\\ 
\label{eqn:Cauchytheta_homo}
\sigma_{\theta \theta} & = P \alpha ( 1 +\alpha^{-3} \rho^2(4 \rho -3) ) +\bigO{\alpha^{-3}}.
\eal
We plot the above Cauchy stress for $\alpha^{-1}$ from $0.05$ to $0.1$, which is within a physiological range, in Figure~\ref{fig:SigmaHomo}. A graph for $\sigma_{rr}/P$ would show all the curves bunched along the same straight line, so instead we have depicted the curves for $\sigma_{rr}/P -\rho$. Figure~\ref{fig:SigmaHomo} reveals that as the relative thickness of the arterial wall increases (shading from blue towards red), the circumferential stress $\sigma_{\theta \theta}$ decreases while the radial stress $\sigma_{rr}$ both increases and becomes less homogeneous.  

\begin{figure}[h!]  
\begin{tikzpicture} 
 	\draw (0,0) node {\includegraphics[width=0.5\textwidth]{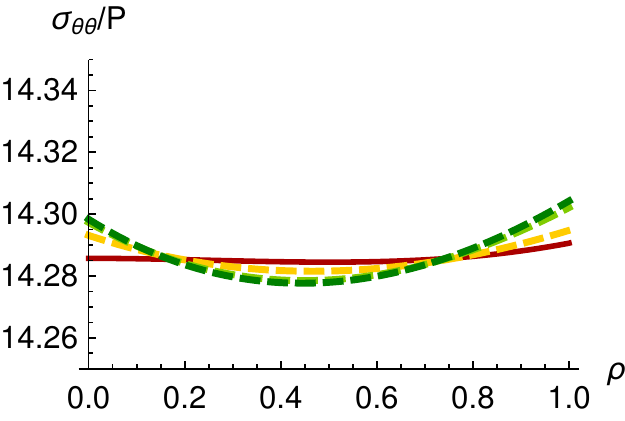}}; 
 	\draw (7.2,0) node { \includegraphics[width=0.5\textwidth]{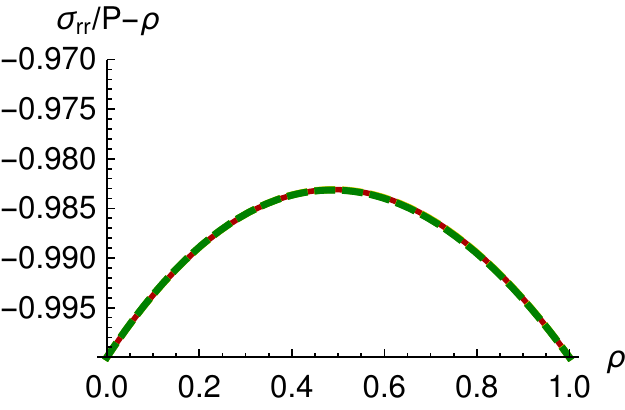}} ;
 	\draw (2.8,.3) node[left,thick]{$ P/\mu$};
	\draw [thick,->] (2.9,-.3) -- (2.9,.3);
\end{tikzpicture}
\caption{The graphs of the dimensionless Cauchy stress against the dimensionless radius $\rho$ for the opening angle method (dashed curves) and the optimal stress~Eqs.~\eqref{eqn:Cauchyr_homo} and~\eqref{eqn:Cauchytheta_homo}  (red solid curves), for which we used parameters for the descending thoracic aorta: $a = \SI{20}{\milli\meter}$, $b= \SI{21.4}{\milli\meter} $, $\alpha^{-1} = 0.07$ and $\lambda_z =1$. As the dashed curves shade from yellow to green 
$P/\mu$ goes through the values $0.05, \, 0.2$ and $0.35$ with   
optimal opening angle $\phi = \SI{65.9}{\degree}, \, \SI{204.2}{\degree}$ and $\SI{261.723}{\degree}$; stress free reference inner radius $\SI{20.3}{\milli\meter}, \, \SI{25.2}{\milli\meter}$ and $\SI{31.1}{\milli\meter}$; and stress free reference outer radius $\SI{22}{\milli\meter}, \, \SI{27.8}{\milli\meter}$ and  $\SI{34.4}{\milli\meter}$ respectively.   
 }
 \label{fig:SigmaCompare}
\end{figure}

A well established method to quantify the residual stresses within arteries is the opening angle method~\parencite{chuong1986residual}. For a neo-Hookean material~\eqref{eqn:PsineoHook} we will use the opening angle method to minimize the circumferential stress component 
\be
\frac{1}{b-a}\int_a^b \left (\frac{d \sigma_{\theta \theta}}{d r}  \right )^2 dr,
\label{int:Drsigmatheta}
\en
in terms of the opening angle $\phi$, restricted to the boundary conditions~\eqref{eqn:sigmaCylindrical_boundarya} and~\eqref{eqn:sigmaCylindrical_boundaryb}. The only two parameters with a unit of time, and mass, are $P$ and $\mu$, so by rewriting $P = P_0 \mu$ the stress will not depend on $\mu$. The results for
%of using the opening angle method to minimize the integral~\eqref{int:Drsigmatheta} with
 $P_0= 0.05$, $P_0= 0.2$ and $P_0= 0.35$ are compared with the optimal stress Eqs.~\eqref{eqn:Cauchyr_homo} and~\eqref{eqn:Cauchytheta_homo} with parameters for the descending thoracic aorta in Figure~\ref{fig:SigmaCompare}. 
   We can see that $\sigma_{\theta  \theta }$ for the opening angle method converges to the optimal stress $\sigma_{\theta \theta}$ as $P/ \mu$ tends to zero, while the plots for $\sigma_{rr}$ all overlap. 
   %Naturally the stress $\sigma_{\theta \theta}$ deceases as $P_0 \to 0$ because when $P_0 =0$ the arterial wall is stress free.
   Note however that $\sigma_{zz}$ for the opening angle method is not necessarily homogeneous in $r$. 
 %while, as discussed at the end of Section~\ref{sec:PlainStrain},  for the constit \eqref{eqn:CauchyInvariants}       
 %is minimized for $\phi \approx \SI{65.9}{\degree}$, $P/\mu \approx 0.05$ (or smaller) and for inner and outer reference radius $20.3$ and $22$ respectively. 
 
We can now invoke Finite Elasticity to derive the residual stress $\vec \tau$ in the unloaded state necessary to sustain the optimal homogeneous Cauchy stress, and then compare our results with the residual stress predicted by the opening angle method. To do so, we let $\Psi$ be a function of both stress and strain and use the initial stress symmetry (ISS), in the form of Eq.~\eqref{eqn:ResidualISS}, to determine $\vec \tau$ as a function of $\vec \sigma$ and $\vec F$. 

\subsection{Residual stress}
\label{sec:Inflation}
%In the previous section we calculated the optimal homogeneous stress gradient without any constitutive assumptions. 
 Here we use Finite Elasticity to connect the current state with the unloaded state.
  %unloaded state parameters with the current state.
   Since the cylindrical symmetry of the artery is maintained when the internal pressure is removed, we describe the unloaded state with the cylindrical coordinates $(R,\Theta, Z)$, then the deformation gradient for the unit basis vectors $\vec e_r$, $\vec e_\theta$, $\vec e_z$, $\vec E_R$, $\vec E_\Theta$ and $\vec E_Z$ becomes 
\be
\label{eqn:InflationDeformation}
\vec F= \frac {\partial r}{ \partial R } \vec e_r \otimes \vec E_R + \frac{r}{R} \vec e_\theta \otimes \vec E_\Theta + \lambda_z \vec e_z \otimes \vec E_Z,
%\begin{pmatrix}
%\frac {\partial r}{ \partial R } & 0 & 0\\
%0 & r/R &0 \\
%0 &0 & \lambda_z
%\end{pmatrix}, 
\en
where we let $\lambda_z$ be a constant. Assuming the material is incompressible we get
\be
 \det \vec F =1 \implies  \frac{\partial r}{ \partial R } \frac{r}{R} \lambda_z = 1.
\en
Let the reference configuration be a hollow cylinder with inner and outer radius $A$ and $B$, respectively, so that  
\be
\label{eqn:r}
r = \sqrt{ (R^2- A^2)\lambda_z^{-1} + a^2 } \implies  \frac{\partial r}{ \partial R } = \frac{R}{r}\lambda_z^{-1},
\en
 which we will use to replace $\partial r / \partial R$ wherever it appears. We will assume that the residual stress $\vec \tau$ is homogeneous in $\Theta$ and $Z$, as the deformation,the Cauchy stress from Eqs.~\eqref{eqn:Cauchyr_homo} and~\eqref{eqn:Cauchytheta_homo}, the boundary conditions~\eqref{eqn:boundary_tau_A} and~\eqref{eqn:boundary_tau_B} are all homogeneous in $\Theta$ and $Z$.  
 
 %To determine the residual stress $\vec \tau$ we use ISS~\eqref{eqn:ResidualISS} applied to Eq.~\eqref{eqn:CauchyInvariants} and substitute $\vec \tau$, $\vec F^{-1}$ and $p_\tau$ respectively in place of $\vec \sigma$, $\vec F$ and $p$. 
 %The Cauchy stress as determined from the previous section is independent of $\theta$ and $z$   
 %The conclusion is that $\vec \tau$ is independent of $\Theta$ and $Z$, $\tau_{ZR} =0$ and $\tau_{Z\Theta} =0$, where we have assumed that $\Psi$ is homogeneous. 
 %Also note that in the previous sections we assumed that $\sigma_{zz}$ is constant, Eq.~\eqref{eqn:CauchyInvariants} shows that $\tau_{ZZ}$ can be chosen to achieve this. 
 
 The equilibrium equations now reduce to
\be
\label{eqn:tauCylindrical}   
   \tau_{\Theta \Theta}= (R \tau_{RR} )' 
\;\; \textrm{ for } A< R <B,
\en
with the zero-traction boundary conditions 
\begin{align}
     \label{eqn:boundary_tau_A}
    & \tau_{RR} = 0 \; \textrm{ for  } R = A,
    \\ 
     & \tau_{RR} = 0 \; \textrm{ for  } R = B.
     \label{eqn:boundary_tau_B}
\end{align}
After making a constitutive choice for $\Psi$, the residual stresses $\tau_{RR}$ and $\tau_{\Theta\Theta}$ will be completely determined from $\vec F$, $\vec \sigma$ and $p_\tau$ due to ISS~\eqref{eqn:ResidualISS}.  

\subsubsection*{$\Psi$ independent of $I_2$, $J_2$, $J_3$ and $J_4$}
To simplify we assume $\Psi$ independent of $I_2$, $J_2$, $J_3$ and $J_4$, so the Cauchy stress~\eqref{eqn:CauchyInvariants} becomes
\be
\vec \sigma = 2 \Psi_{I_1} \vec F \vec F^{T} - p \vec I + 2 \Psi_{J_1} \vec F \vec \tau \vec F^T,
\en  
%\vec \sigma = (\alpha + \beta \tr \vec \tau) \vec F \vec F^{T} - p \vec I + \vec F \vec \tau \vec F^T,
and, from ISS~\eqref{cond:ISS}, we can swap $\vec \tau$, $\vec F$ and $p$ respectively with $\vec \sigma$, $\vec F^{-1}$ and $p_\tau$ to get
\be
%\vec \tau = (\alpha + \beta \tr \vec \sigma) \vec F^{-1} \vec F^{-T} - p_\tau \vec I + \vec F^{-1} \vec \sigma \vec F^{-T},
\vec \tau = 2 \Psi_{I_1}^\sigma \vec F^{-1} \vec F^{-T} - p_\tau \vec I + 2 \Psi_{J_1}^\sigma \vec F^{-1} \vec \sigma \vec F^{-T}.
\en  
  Substituting $\vec F$ from Eq.~\eqref{eqn:InflationDeformation} into the above equation we arrive at
\bal
\label{eqn:tauR}
\tau_{RR} &= 2  \lambda_z^2 \frac{r^2}{R^2} \left (\Psi_{I_1}^\sigma + \Psi_{J_1}^\sigma  \sigma_{rr}  \right ) - p_\tau ,
\\
\label{eqn:tauTheta}
\tau_{\Theta \Theta} &= 2 \frac{R^2}{r^2} (\Psi_{I_1}^\sigma +  \Psi_{J_1}^\sigma  \sigma_{\theta \theta} ) - p_\tau.
%\\
%\label{eqn:tauZ}
%\tau_{ZZ} &=  2 \lambda_z^{-2} \left( \Psi_{I_1}^\sigma + \Psi_{J_1}^\sigma \sigma_{z z} \right ) - p_\tau.
\eal
%\bal
%\tau_{RR} &= (\alpha + \beta (\sigma_{rr}+\sigma_{\theta \theta}+\sigma_{zz}) ) \lambda_z^2 \frac{r^2}{R^2} - p_\tau + \lambda_z^{-2} \frac{R^2}{r^2} \sigma_{rr},
%\tau_{\Theta \Theta} = (\alpha + \beta \tr \vec \sigma) \frac{R^2}{r^2} - p_\tau \vec I + \vec F^{-1} \vec \sigma \vec F^{-T},
%\eal
%ISS further requires that $\Psi$, and possibly $p$ and $p_\tau$, be choosen so that the Eqs.~\eqref{eqns:ISSJ1} hold.
%Let us now consider equilibrium~\eqref{eqn:tauCylindrical}. 
 We remark that $\tau_{ZZ}$ has not been derived since we have not taken into consideration $\sigma_{zz}$ in Section~\ref{sec:MinimalStressGrad}.
  
  %However, ISS further imposes Eqs.~\eqref{eqns:ISSJ1}${}_{3,4}$, which depending on our choice for $\Psi$, may determine $p$ and $p_\tau$ in terms of $\Psi$. 
 %To leave $p_\tau$ undetermined we adopt the neo-Hookean material model~\eqref{eqn:Psi2D}, which results in $\Psi_{J_1}^\sigma =1/2$ and $2 \Psi_{I_1}^\sigma = \ini p(\vec \sigma)$ where $\ini p$ is given by substituting $\vec \tau$ for $\vec \sigma$ restricted to $(r, \theta)$ in $\ini p_{+}$ of Eq.~\eqref{eqn:p2D}. 
 %For the numerical results ahead we found that $-$ in place of $\pm$ in Eq.~\eqref{eqn:p2D} gave unphysical results. 
 %As we will use Eqs.~\eqref{eqn:Cauchyr_homo} and~\eqref{eqn:Cauchytheta_homo}, 
To compare unloaded arteries of different sizes we write $r$ and $R$ in terms of the dimensionless radius $\rho$,
\be
r = (b- a)\rho +a \;\; \text{ and } \;\; R = \sqrt{A^2 + \lambda_z \left [ (a + (b-a) \rho)^2 - a^2  \right ] },  
\en
which we use to rewrite the
 %the left handside of 
ODE~\eqref{eqn:tauCylindrical} in the form
\bal
    & R \frac{d \tau_{RR}}{d R} + \tau_{RR} - \tau_{\Theta \Theta} = R \frac{d \tau_{RR}}{d \rho} \frac{d \rho}{d R} + \tau_{RR} - \tau_{\Theta \Theta} 
    \\
     &=    \frac{d \tau_{RR}}{d \rho} R \left ( \frac{d R}{d \rho}  \right )^{-1}  + \tau_{RR} - \tau_{\Theta \Theta} =0,
\eal
which implies that
\be
     \frac{d \tau_{RR}}{d \rho}    + \frac{1}{R} \frac{d R}{d \rho} \left ( \tau_{RR} - \tau_{\Theta \Theta} \right) =0. 
\en
%After substituting~\eqref{eqn:tauR} and~\eqref{eqn:tauTheta}, the right handside of the above Eq. is independent of $p_\tau$, thus we can 
Integrating both sides in $\rho$ and using the unloaded boundary condition~\eqref{eqn:boundary_tau_A} and~\eqref{eqn:boundary_tau_B}  we reach
\be
 \label{ode:ptauIntegrated}
    \tau_{RR} =  \int_0^R \frac{\tau_{\Theta \Theta} - \tau_{RR}}{R} \frac{d R}{d \rho}d \rho=0 \;\; \text{and} \;\;  \int_0^1 \frac{\tau_{\Theta \Theta} - \tau_{RR}}{R} \frac{d R}{d \rho}d \rho=0.
\en 
%Evaluating the above for $R=B$ and using the boundary condition~\eqref{eqn:boundary_tau_B} we reach an integral equation, 
Eq.~\eqref{ode:ptauIntegrated}${}_2$ is independent of $p_\tau$, and can be used to solve for one of the parameters $P$, $a$, $b$, $A$, $\lambda_z$ or $\mu$, note that $B$ depends on the other parameters because Eq.~\eqref{eqn:r} evaluated at $R=B$ gives $B^2=  A^2 +\lambda_z (b^2- a^2) $. The only two parameters with a unit of time or mass are $P$ and $\mu$, so by rewriting $P = P_0 \mu$ the Eqs.\eqref{ode:ptauIntegrated} will not depend on $\mu$.
To illustrate, we adopt the neo-Hookean material model~\eqref{eqn:Psi2D}, which results in $\Psi_{J_1}^\sigma =1/2$ and $2 \Psi_{I_1}^\sigma = \ini p(\vec \sigma)$ where $\ini p$ is given by substituting $\vec \tau$ for $\vec \sigma$ restricted to $(r, \theta)$ in $\ini p_{+}$ of Eq.~\eqref{eqn:p2D}. 

\begin{figure}
\centering
\includegraphics[width=0.9\textwidth]{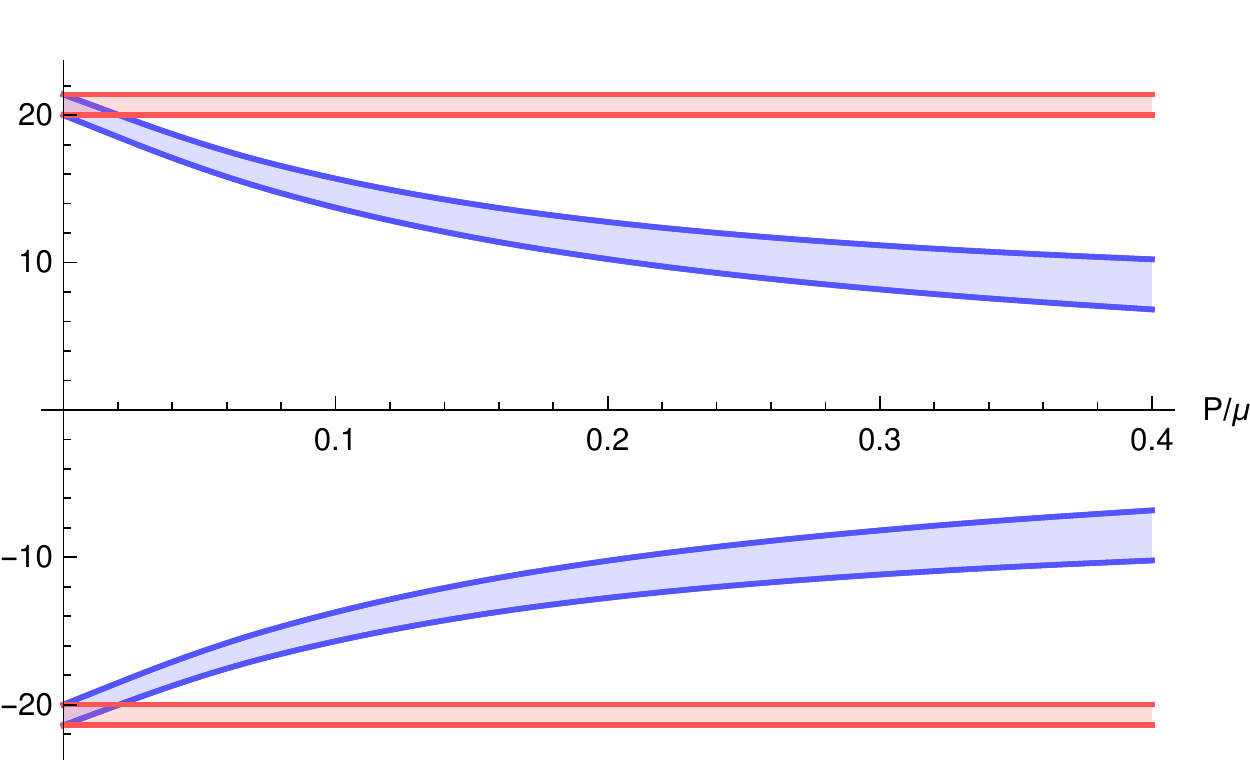}
\caption{The loaded artery wall is illustrated by the red curves, while the unloaded artery wall is illustrated by the blue curves. The $y$--axis illustrates the position of the loaded and unloaded walls measured from the center of the artery.  
  The stress in the loaded geometry is taken to be the optimal Cauchy stress~\eqref{eqn:Cauchyr_homo} and~\eqref{eqn:Cauchytheta_homo}, and we assume a residually stressed neo-Hookean material~\eqref{eqn:Psi2D}. 
  %The unloaded geometry for the opening angle method is approximately the same.
  }
\label{fig:UnloadedAndLoaded}
\end{figure}

To calculate the residual stress that supports the optimal Cauchy stress, we substitute $\sigma_{\theta \theta}$ and $\sigma_{r r}$ Eqs.~\eqref{eqn:Cauchyr_homo} and~\eqref{eqn:Cauchytheta_homo} into $\tau_{\Theta \Theta}$ and $\tau_{RR}$ Eqs.~\eqref{eqn:tauR} and~\eqref{eqn:tauTheta}. We can then use Eq.~\eqref{ode:ptauIntegrated}${}_2$ to determine the unloaded geometry from the loaded geometry. To illustrate we take the parameters for the descending thoracic aorta, as used in the previous section, $a=\SI{20}{\milli\meter}$, $b=\SI{1.4}{\milli\meter}$, and $\lambda_z=1$. We can then determine the unloaded inner radius $A$ for different values of $P_0$, which is shown in Figure~\ref{fig:UnloadedAndLoaded}. Surprisingly the unloaded geometry given by the opening angle method is approximately the same as shown in Figure~\ref{fig:UnloadedAndLoaded}.   
 %Using $\tau_{RR}$~\eqref{eqn:tauR} for the neo-Hookean material we can write the unloaded boundary conditions~\eqref{eqn:boundary_tau_A} and~\eqref{eqn:boundary_tau_B} as
%\bal
%\label{eqn:UnloadedHomoA}
%& p_\tau(0) = \lambda_z^2 \frac{a^2}{A^2} \left( \ini p(\vec \sigma) -  P \right )  \;\; \text{for} \;\; \rho=0,
%\\
%\label{eqn:UnloadedHomoB}
%& p_\tau(1) = \lambda_z^2 \frac{b^2}{B^2} \ini p(\vec \sigma) \quad \quad \quad  \quad  \text{for} \;\; \rho=1,
%\eal 
%where we have used that $\sigma_{rr} = -P$ at $\rho=0$ and $\sigma_{rr} = 0$ at $\rho=1$. 

%If we assume $a$ and $b-a$ are given from in-vivo measurements, then we are left with three unknown parameters $P_0$, $A$ and $B$, one of which can be determined through~\eqref{ode:ptauIntegrated}.

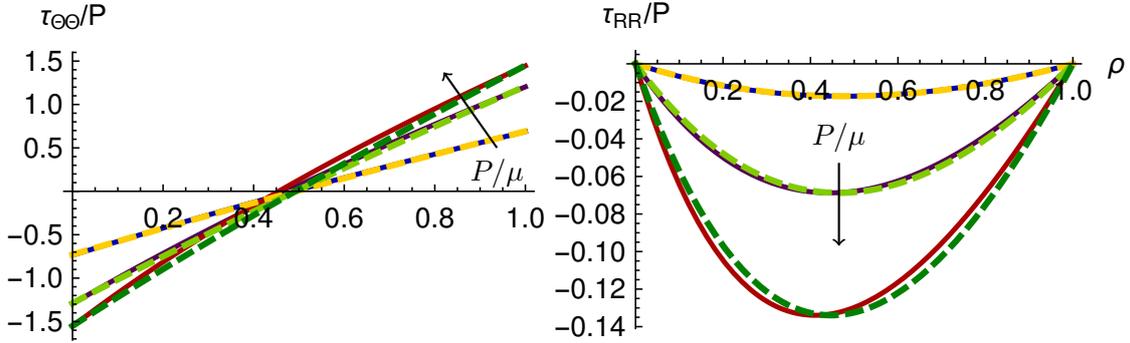
\begin{figure}[h!]  
\begin{tikzpicture} 
 	\draw (0,0) node {\includegraphics[width=0.5\textwidth]{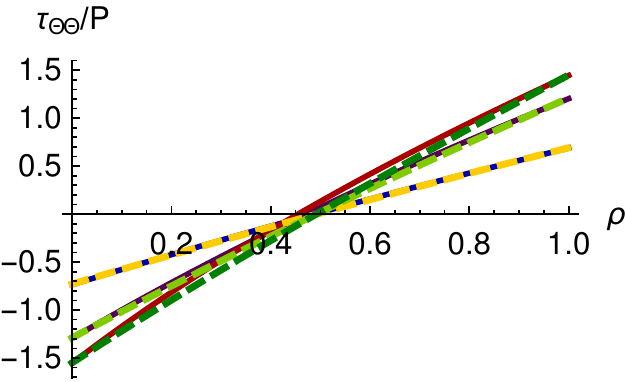}}; 
 	\draw (7.2,0) node {\includegraphics[width=0.5\textwidth]{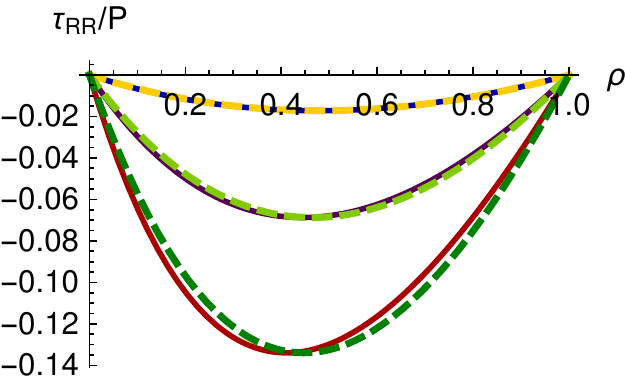}} ;
 	\draw (2.7,.3) node[below,thick]{$ P/\mu$};
	\draw [thick,->] (2.7,.3) -- (2.,1.3);
 	\draw (7.2,.1) node[above,thick]{$ P/\mu$};
	\draw [thick,->] (7.2,.1) -- (7.2,-1.);
\end{tikzpicture}

\caption{shows the dimensionless residual stress against the dimensionless radius $\rho$. The dashed curves are from the opening angle method and the solid curves are from Eq.~\eqref{eqn:tauCylindrical} together with the optimal stresses~Eqs.~\eqref{eqn:Cauchyr_homo} and~\eqref{eqn:Cauchytheta_homo}. As $P/\mu$ goes through the values $0.05, \, 0.2$ and $0.35$, the dashed curves shade from yellow to green and the solid curves shade from blue to red. The unloaded geometry for both methods is approximately the same and shown in Figure~\ref{fig:UnloadedAndLoaded}.
   %with unloaded inner and outer radius  $ A = \SI{16.4}{\milli\meter}$ and $B \approx \SI{18.1}{\milli\meter}$ respectively. The blue and green curves have $P/\mu = 0.2$ with unloaded inner and outer radius $A \approx \SI{10.2}{\milli\meter}$ and $B \approx \SI{12.8}{\milli\meter} $ respectively. 
   The other parameters are $a = \SI{20}{\milli\meter}$, $b= \SI{21.4}{\milli\meter} $, $\alpha^{-1} = 0.07$ and $\lambda_z =1$. 
 }
 \label{fig:TauCompare}
\end{figure}

We can also investigate the residual stress in the unloaded state. Once $A$ is determined from Eq.~\eqref{ode:ptauIntegrated}${}_2$ we can use Eq.~\eqref{ode:ptauIntegrated}${}_1$ to determine $\tau_{RR}$, and then $\tau_{\Theta \Theta}$ from Eq.\eqref{eqn:tauCylindrical}. To compare with the opening angle method we choose $P/\mu = 0.05, \, 0.2$ and $0.35$, which all give a reasonable unloaded geometry, see Figure~\ref{fig:UnloadedAndLoaded}. The results are shown in Figure~\ref{fig:TauCompare}.  

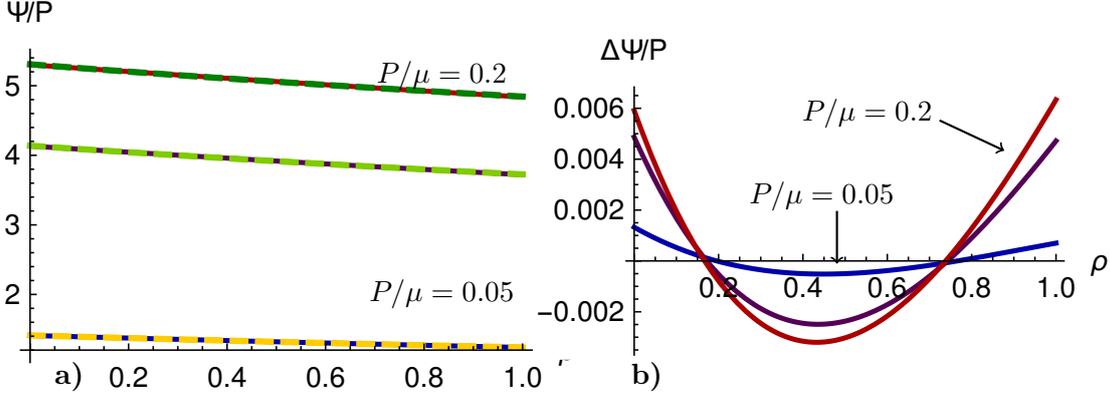
\begin{figure}[h!]  
\begin{tikzpicture} 
 	\draw (0,0) node {\includegraphics[width=0.5\textwidth]{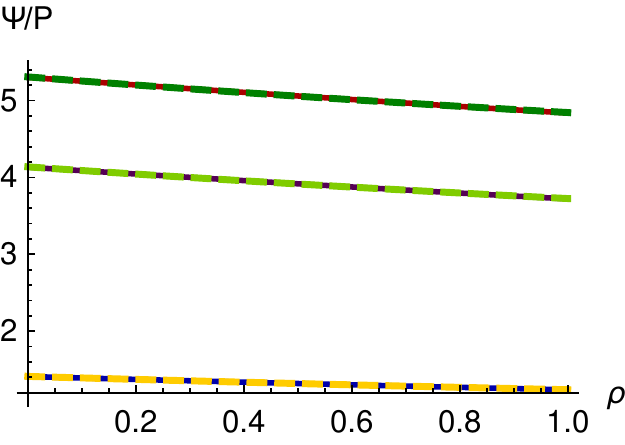} }; 
 	\draw (7,0) node {\includegraphics[width=0.5\textwidth]{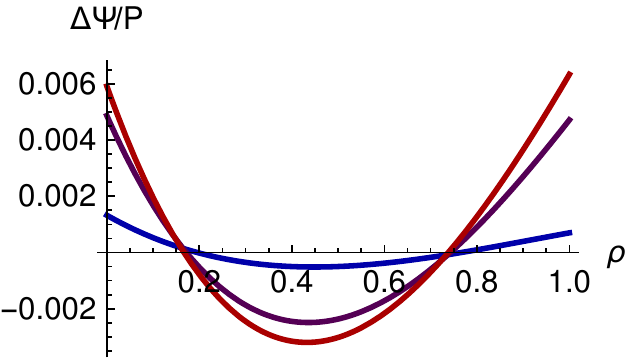}} ;
 	\draw (2.,1.6) node[thick]{$ P/\mu = 0.2$};
 	\draw (2.,-1.3) node[thick]{$ P/\mu = 0.05$};
 	\draw (7.6,1.1) node[thick]{$ P/\mu = 0.2$};
 	\draw (7.,0.) node[thick]{$ P/\mu = 0.05$};
 	\draw (-2.9,-2.4) node[thick]{\textbf {a)}};
 	\draw (-2.8+7.5,-2.4) node[thick]{\textbf {b)}};
	\draw [thick,->] (8.55,1.) -- (9.4,.6);
	\draw [thick,->] (7.2,-0.2) -- (7.2,-.9);
\end{tikzpicture}
\caption{\textbf{(a)} compares the dimensionless strain energy density $\Psi/P$ versus the dimensionless radius $\rho$ resulting from the opening angle method, yellow (green) dashed curves, with the model \eqref{eqn:Psi2D} whose residual stress maintains the optimal stress~Eqs.~\eqref{eqn:Cauchyr_homo} and~\eqref{eqn:Cauchytheta_homo}, the red (blue) solid curves. The parameters used are given in Figure~\ref{fig:TauCompare}. \textbf{(b)} shows $\Psi/P$ from the opening angle method minus $\Psi/P$ from the model \eqref{eqn:Psi2D}, which we denote as $\Delta \Psi/P$.
 %The red and yellow curves have $P/\mu = 0.05$, with unloaded inner and outer radius $ A \approx \SI{16.4}{\milli\meter}$ and $B \approx \SI{18.1}{\milli\meter}$ respectively. The blue and green curves have $P/\mu = 0.2$ with unloaded inner and outer radius $A \approx \SI{10.2}{\milli\meter}$ and $B \approx \SI{12.8}{\milli\meter} $ respectively. The other parameters are $a = \SI{20}{\milli\meter}$, $b= \SI{21.4}{\milli\meter} $, $\alpha^{-1} = 0.07$ and $\lambda_z =1$. 
 }
 \label{fig:PsiCompare}
\end{figure}

Biological tissues are very energy efficient. It is a common line of reasoning that biological systems adapt so as to minimize their potential energy. 
So it is possible that biological materials remodel their residual stress to lower their potential energy.
%So it is possible that residual stress is used to minimize the free energy density $\Psi$. 
Figure~\ref{fig:PsiCompare}a below shows the free energy density $\Psi$ for the opening angle method and for the model \eqref{eqn:Psi2D} with the optimal stresses~Eqs.~\eqref{eqn:Cauchyr_homo} and~\eqref{eqn:Cauchytheta_homo}. In all cases $\Psi$ is approximately a straight line that descreases as $\rho$ increases away from the pressured boundary $\rho=0$ ($r=a$). Figure~\ref{fig:PsiCompare}b shows the difference between $\Psi/P$ from the opening angle method minus $\Psi/P$ from the model \eqref{eqn:Psi2D}, which we denote as $\Delta \Psi/P$. The integral of $\Delta \Psi/P$ over $\rho \in [0,1]$ is positive if the opening angle method has on average a larger free energy density than the model \eqref{eqn:Psi2D}, with the optimal stresses Eqs.~\eqref{eqn:Cauchyr_homo} and~\eqref{eqn:Cauchytheta_homo}. We found that as $P/\mu$ increased, so did the total free energy in the cylinder cross-section $\int_0^1\Delta \Psi/P (b-a)(\rho (b-a) +a ) d\rho$; for instance, this integral evaluates to $-1.64 10^{-4} $ for $P/\mu = 0.05$, $2.37 10^{-3}$ for $P/\mu =0.2$ and $4.25 10^{-3}$ for $P/\mu =0.35$. 

Essentially we can see from the Figures~\ref{fig:SigmaCompare} and~\ref{fig:PsiCompare} that the optimal stresses~Eqs.~\eqref{eqn:Cauchyr_homo} and~\eqref{eqn:Cauchytheta_homo} with the neo-Hookean material~\eqref{eqn:Psi2D} produce a more homogeneous stress and lower free energy than the opening angle method, though the two methods gave very similar results. The advantage in using the ISS~\eqref{cond:ISS} for a homoeostasis hypothesis for the Cauchy stress $\vec \sigma$ is two fold. First is the freedom to choose $\vec \sigma$ so as to satisfy the homoeostasis hypothesis, which with ISS becomes a separate step from choosing a constitutive equation. Second is that it can be relatively straightforward to use ISS~\eqref{cond:ISS} to quantify the residual stress needed to support the chosen  $\vec \sigma$.          
%As $P/\mu \to 0$ the loaded configuration becomes stress free, and for this reason $\Delta \Psi/P \to 0$ as $P/\mu \to 0$.

\section{Conclusions}

In order to quantify the initial stress within a solid using non-destructive experimental techniques, it is  convenient to write the free energy density $\Psi= \Psi(\vec F, \vec \tau)$ in terms of the deformation gradient $\vec F$ and the initial stress $\vec \tau$.

 %For biological tissues the initial stress can be due to a number of factors, such as growth, remodelling, external loads and active contraction. So developing models of the free energy in terms of internal stress, irrespective of the origins of that stress, may simplify existing models and lead to new modelling techniques for stressed biological systems. 

%One such example was given in Section~\ref{sec:HomogeneousStress}.
%For situations where the initial stress $\vec \tau$ can be measured or determined a priori, it would be ideal to have
%the free energy density $\Psi= \Psi(\vec F, \vec \tau)$ written explicitly in terms of the deformation gradient $\vec F$ and initial stress $\vec \tau$. Models for $\Psi= \Psi(\vec F, \vec \tau)$ where 
%For example, calculating $\vec \tau$ is often simpler than calculating a finite deformation from a stress-free configuration that gives rise to the initial stress. 

In this article we presented a new constitutive condition, the initial stress symmetry, that aids
in proposing suitable constitutive relations for $\Psi = \Psi(\vec F, \vec \tau)$ by providing nine scalar equations, see Eqs.~\eqref{eqn:ISS1} and~\eqref{eqn:ISS2}. One immediate result is that guessing a functional dependence for $\Psi = \Psi(\vec F, \vec \tau)$ is not a trivial task. In fact all choices for $\Psi(\vec F, \vec \tau)$ used in the literature do not satisfy ISS. Conversely, using ISS, we proposed two simple choices for $\Psi =\Psi(\vec F, \vec \tau)$, one incompressible Eq.~\eqref{eqn:PsiInTau} and one compressible Eq.~\eqref{eqn:PsiCompressibleNeoHook}, which include a generalization for an initially stressed neo-Hookean material.

One consequence of ISS is that the initial stress can be derived from the Cauchy stress. Furthermore, ISS suggests that it is possible to first choose the Cauchy stress and, second, to make a constitutive choice and then use Eq.~\eqref{eqn:ResidualISS} to determine the corresponding initial stress. We illustrated this method in Section~\ref{sec:HomogeneousStress} for a simplified arterial wall, where we chose the ideal Cauchy stress by using a minimal stress gradient hypothesis as the homeostatic condition, and then we calculated the optimal residual stresses of the unloaded remodelled artery.

Since initial stresses are widespread in both inert and living matter, the proposed constitutive theory can be used in many applications towards a non-destructive quantification of initial tensions in solids. Future research include determining of a wider class of material behaviours, e.g. including a natural material anisotropy, and linking the initial stress to elastic wave speeds.

%Is the initial stress symmetry important? If the free energy is a function of solely $X$, $\vec F$ and $\vec \tau$, then to not accept ISS is to... 

%\emph{This section is really a pre-draft and will be mostly re-written. }

%Point out what were the major contributions of the article, what is useful here? Tell the whole story of the article and point to where forward.

%Why is the initial stress symmetry important? Returning to Figure~\ref{fig:Configurations}, to illustrate suppose that the stress $\vec \tau \not = \hat {\vec \varsigma }(\vec F^{-1}, \vec \sigma, p_\tau)$. This would imply that the constitutive Eq.~\eqref{eqn:CauchyStressGeneral} does not hold when $\bb$ is the reference configuration, which in turn means that either $\Psi$ does not depend on solely $X$, $\vec F$ and $\vec \tau$, or alternatively, one of the assumptions used to deduce the constitutive Eq.~\eqref{eqn:CauchyStressGeneral} in elasticity is incorrect~\parencite{Guillou2006Residual}. 

%However there are many cases in which it does seem natural to have the free energy density as a function of solely $X$,$\vec F$ and $\vec \tau$, and, the typical assumptions used in elasticity seem very reasonable~\parencite{marsden1994mathematical}.  

%%%%%%%%%%%%%%%%%%%%%%%%%%%%%%%%%%
\section*{Acknowledgements}
Partial funding by the Irish Research Council, the Royal Society and the Hardiman Scholarship programme at the National University of Ireland Galway, are gratefully acknowledged. 

%%%%%%%%%%%%%%%%%%%%%%
\printbibliography

%\appendixpageoff

\begin{appendices}

\renewcommand{\theequation}{\Alph{section}.\arabic{equation}}
\setcounter{equation}{0}

\section{Stress-free configuration implies ISS}
\label{app:VirtualConfig}
Referring to Figure~\ref{fig:Configurations}, the virtual stress-free configuration guarantees that
for any $\vec \tau$ and $\vec F$ there exists a $\ini {\vec F}$ such that
%\be
 %\label{eqn:SigmaVirtual}
%\vec \sigma = \hat{\vec \varsigma} (\tilde {\vec F} ,\vec 0, \hat p(\vec \sigma, \tilde{\vec F}, 0) )  \quad \text{and} \quad \vec \tau = \hat {\vec \varsigma}( \ini{\vec F},\vec 0, \hat p(\vec \tau, \ini{\vec F},0) ).
%\en
\be
 \label{eqn:SigmaVirtual}
\vec \sigma = \hat{\vec \vartheta} ( \tilde {\vec B} ,  p  )  \quad \text{and} \quad \vec \tau = \hat {\vec \vartheta}( \ini{\vec B}, \ini p).
\en
 where 
 %$\hat{\vec \vartheta}$ is a constitutive choice
  %that describes the elastic response of the material;
   $\ini{\vec B} = \vec F^{-1} \tilde {\vec B} \vec F^{-T}$; $p$ and $\ini p$ are a scalar fields where $p$ depends on the boundary conditions on $\bb$ and $\tilde {\vec B}$, while $\ini p$ depends on the boundary conditions on $\ini \bb$ and $\ini {\vec B}$;  and
 \[
 \hat{\vec \vartheta} (  {\vec A \vec A^T} ,  q  ) \coloneqq \vec A \frac{\partial \Psi}{\partial (\vec A^T \vec A) } \vec A^T - q \vec I. 
 %\quad \text{for every } \vec A \in \mathbb R^{3\times 3} 
 \]
 By setting $\vec \tau =0$ in Eq.~\eqref{eqn:CauchyInvariants} we can see that the right side depends only on $\vec A \vec A^T$ and $q$.
 
 \cite{johnson1995use} demonstrated that it is possible to determine $\ini{\vec B}$ from any given $\vec \tau$, together with appropriate boundary conditions on $\ini \bb$, in order to reach useful constitutive equations such as
  %such that  $\vec \sigma = \hat{\vec \varsigma} ( {\vec F} ,\vec \tau, p )$ 
 %Using equation~\eqref{eqn:CauchyStressGeneral} with the above we have that
 %The two forms of calculating $\vec \sigma$, i.e. $\hat{\vec \varsigma} (\tilde {\vec F} ,\vec 0, p )$ and $\hat{\vec \varsigma} ( {\vec F} ,\vec \tau, p ) $, can be shown to be equivalent through the use of thermodynamics~\parencite{guillou2006growth}, where $\tilde {\vec F} =  {\vec F} \ini {\vec F}$. To summarize: 
  %However we can illustrate their equivalence by considering that $\tilde {\vec F}_1$ is a dependant variable completely determined by $\vec F_1$, then in component form
%\begin{multline} 
%\tilde { F}_{ab} \frac{\partial \Psi}{\partial \tilde{ F}_{c b} } =  { F}_{ad} \ini { F}_{db}  \frac{\partial \Psi}{\partial { F}_{i j} } \frac{\partial F_{ij}}{\partial \tilde{ F}_{c b} } = 
%{ F}_{ad} \ini { F}_{db}  \frac{\partial \Psi}{\partial { F}_{i j} } \frac{\partial \tilde F_{ik} {\ini F^{-1}_{kj}} }{\partial \tilde{ F}_{c b} }  =\\
 %{ F}_{ad} \ini { F}_{db}  \frac{\partial \Psi}{\partial { F}_{c j} }  {\ini F^{-1}_{bj}} = { F}_{ab}   \frac{\partial \Psi}{\partial { F}_{c j} },
%\end{multline}
%where we have omitted the argument of $\Psi$. 
 %\be
 %\label{eqn:StressEquivalence}
 %\vec \sigma =\hat{\vec \varsigma}( \vec F \ini{\vec F} , \vec 0, \hat p(\vec \sigma, \vec F \ini{\vec F}, 0) ) = \hat{\vec \varsigma }(\vec F,\vec \tau, \hat p(\vec \sigma, \vec F,  \vec \tau)) \quad \text{for any} \; \vec F \; \text{and} \; \ini{\vec F},
 %\en
 \be
 \label{eqn:StressEquivalence}
 %\vec \sigma =
 \hat{\vec \vartheta}( \vec F \ini{\vec B} \vec F^T ,  p ) = \hat{\vec \varsigma }(\vec F,\vec \tau, p) \quad \text{for any} \; \vec F \; \text{and} \; \ini{\vec B}.
 \en
 %where now $p$ is a function of $\vec F$, $\vec \tau$ and the boundary conditions on $\bb$.
  %In section~\ref{sec:SimpleEnergy} we see that for the neo-Hookean material there are three possible choices for $\hat{ \vec \varsigma}$,  see equations~\eqref{eqn:tau} and~\eqref{eqn:inip}.
  %(for certain values of $\vec \tau$)
   %One of these choices for $\hat{ \vec \varsigma}$ will give the type of virtual configuration described in~ \cite{johnson1995use}. 
    %Depending on our choice, equation~\eqref{eqn:StressEquivalence} may not hold for every $\vec F$ and every $\vec \tau$ (or likewise $\ini{ \vec B}$).  
  %$\ini {\vec B}$ is not uniquely determined by $\vec \tau$, see equations~\eqref{eqn:tau} and~\eqref{eqn:inip}. 
 %It is possible however to choose $\hat{ \vec \varsigma}$, the case~\eqref{eqn:inip}${}_1$, so that equation~\eqref{eqn:StressEquivalence} does hold.
 
 %(as the other two cases cease to exist for specific regions for $\vec \tau$)
  As Eq.~\eqref{eqn:SigmaVirtual}${}_2$ is valid for any $\vec F$ and $\ini {\vec B}$, we can substitute
  \be
  \vec F \text{ for } \vec F^{-1}, \quad \ini{\vec B} \text{ for } \tilde{\vec B} 
  \label{eqns:swap}
  \en
    and swap the boundary conditions on $\bb$ and $\ini \bb$ (these substitutions effectively swap $\bb$ for $\ini \bb$), so that Eq.~\eqref{eqn:SigmaVirtual}${}_2$ becomes $\vec \sigma = \hat{\vec \vartheta}(\tilde{\vec B},  p)$, 
    %\be 
 %\label{eqn:tauVirtual2}
 %\vec \sigma = \hat{\vec \vartheta}(\tilde{\vec B},  p),
 %\en
 where we have assumed that the above together with the boundary conditions on $\bb$ has $\vec \sigma$ as the unique solution. Analogously, Eq.~\eqref{eqn:SigmaVirtual}${}_1$ becomes $\vec \tau = \hat {\vec \vartheta}( \ini{\vec B}, \ini p)$.
 
 This means that when we make the substitutions~\eqref{eqns:swap} in Eq.~\eqref{eqn:StressEquivalence} we should also swap $\vec \tau$ and $\vec \sigma$, so that
  \be
 \label{eqn:StressEquivalence2}
 \hat{\vec \vartheta}( \ini {\vec B},  \ini p) = \hat{\vec \varsigma }(\vec F^{-1},\vec \sigma, \ini p), 
 \en
 %There is a problem here, why does $\hat p( ...)$ get swapped for $\ini p$? Without the explicit arguments this seems unclear.
   The left-hand side is simply $\vec \tau$ and the right-hand side is $\vec \tau$ given by ISS~\eqref{cond:ISS}. Finally, the above condition is identically true, since this result is valid for every $\vec F$ and $\ini{\vec B}$. Moreover, we assumed that 
   %the boundary conditions together with equation~\eqref{eqn:tauVirtual} uniquely specifies
    $\ini{\vec B}$ can be determined from $\vec \tau$, so Eq.~\eqref{eqn:StressEquivalence2} holds for every $\vec F$ and $\vec \tau$.
   
    %this assumption is often used~\parencite{Hoger1997}. So given our assumptions, we may invert $\ini{\vec F}$ for $\vec \tau$ and state that $\vec \tau = \hat{\vec \varsigma }(\vec F^{-1},\vec \sigma, \hat p(\vec \tau, \vec F^{-1},  \vec \sigma))$ for every $\vec F$ and $\vec \tau$, as long as $\vec \sigma = \hat{\vec \varsigma }(\vec F,\vec \tau, \hat p(\vec \sigma, \vec F,  \vec \tau))$ which is exactly the ISS condition~~\eqref{cond:ISS}.

\section{Reducing ISS to nine scalar equations}
\label{app:ReduceISS}

Deriving the scalar equations equivalent to ISS requires lengthy calculations. First we use the stress expansion~\eqref{eqn:CauchyInvariants} and ISS~\eqref{cond:ISS} to write the residual stress as
\begin{multline}
\vec \tau = -p_{\vec \tau} {\vec I} + 2  \Psi_ {I_1}^{\vec \sigma} {\vec C^{-1}}+2\Psi^{\vec \sigma}_{I_2}( I_2  {\vec C^{-1}}-{\vec C^{-2}})+2 \Psi_{J_1}^{\vec \sigma} {\vec F^{-1} \vec \sigma \vec F^{-T}}+2\Psi_{J_3}^{\vec \sigma} \vec F^{-1} \vec \sigma^2 \vec F^{-T} 
\\ 
  +2\Psi_{J_2}^{\vec \sigma}\vec F^{-1}( \vec \sigma {\vec B^{-1}}+\vec B^{-1} \vec \sigma ) \vec F^{-T} + 2 \Psi_{J_4}^{\vec \sigma} \vec F^{-1}( \vec \sigma^2 \vec B^{-1}+\vec B^{-1} \vec \sigma^2 ) \vec F^{-T}
\label{eqns:TauStressInvariants}
\end{multline}
where $I_2 = \tr ({\vec C^{-1}})$ from Cayley-Hamilton with $\det \vec C =1$. The above, together with~\eqref{eqn:CauchyInvariants} for $\vec \sigma$, must hold for every $\vec \tau$ and $\vec C$. To make use of this restriction, we substitute $\vec \sigma$ from~\eqref{eqn:CauchyInvariants} into Eq.~\eqref{eqns:TauStressInvariants} to write an equation in terms of only $\vec \tau$, $\vec C$, $p$ and $p_{\vec \tau}$, compactly written as
\be
\alpha_{ijkmn}\vec C^i \vec \tau^j \vec C^k \vec \tau^m \vec C^n =0,
\label{eqn:tauCoaxialC}
\en
where we adopt the convention that repeated indices in a single term implies summation over all the values of the index, with $i, k, n \in \{-2,-1, 0,1,2\}$ and $j,m \in \{0,1,2\}$. The $\alpha_{ijk}$'s can be calculated with a computer algebra system and are too cumbersome to reproduce here, though it is important to remember that the $\alpha_{ijk}$'s are functions of $p$ and $p_{\vec \tau}$; the invariants \eqref{eqns:Is}, \eqref{eqns:Itau} and \eqref{eqns:Icombined}; and the derivatives of $\Psi$ with respect to the invariants~\eqref{eqns:Is} and \eqref{eqns:Icombined}. Due to the Eqs.~\eqref{eqns:TauStressInvariants} and~\eqref{eqn:CauchyInvariants} being symmetric, we have the symmetries $\alpha_{ijkmn} = \alpha_{njkmi} = \alpha_{nmkji}$.

Our aim here is to use Eq.~\eqref{eqn:tauCoaxialC} to obtain the corresponding scalar equations that hold for any choice of $\Psi$ and the invariants \eqref{eqns:Is}, \eqref{eqns:Itau} and \eqref{eqns:Icombined}. To acheive this, we need to understand how we can vary $\vec C$ and $\vec \tau$ while keeping the invariants fixed. We can then apply these variations to Eq.~\eqref{eqn:tauCoaxialC}, which holds true for any $\vec C$ and $\vec \tau$, to reach scalar equations in terms of the invariants and $\Psi$.

So first, to keep the invariants~\eqref{eqns:Is} and \eqref{eqns:Itau} fixed, we have to keep the eigenvalues of both $\vec C$ and $\vec \tau$ fixed. We can assume that the eigenvalues $\lambda_1^2$, $\lambda_2^2$ and $\lambda_3^2$ of $\vec C$ are all different, as this is true for most any deformation and so we do not loose generality. The same assumption can be made for $\tau_1$, $\tau_2$ and $\tau_3$, the eigenvalues of $\vec \tau$. 

The mixed invariants~\eqref{eqns:Icombined} involve the eigenvectors of $\vec C$ and $\vec \tau$. So we write the eigen decompositions
\be
\vec C = \sum_{i=1}^3 \lambda_i^2 \vec V_i \otimes \vec V_i \quad \text{and} \quad \vec \tau = \sum_{j=1}^3 \tau_j \vec t_j \otimes \vec t_j, 
\en
where the $\vec V_i$'s and $\vec t_j$'s are respectively the eigenvectors of $\vec C$ and $\vec \tau$. Using these decompositions we see that
\bal
\label{eqns:ConstantMixedInvariants1}
& J_1 = \tr (\vec \tau \vec C) =  \tau_j \lambda_i^2 (\vec V_i \cdot \vec t_j)^2 
, \; J_2 = \tr (\vec \tau \vec C^2) = \tau_j \lambda_i^4 (\vec V_i \cdot \vec t_j)^2, 
\\
\label{eqns:ConstantMixedInvariants2}
& J_3 = \tr ( \vec \tau^2 \vec C) =  \tau_j^2 \lambda_i^2 (\vec V_i \cdot \vec t_j)^2 
, \; J_4 = \tr ( \vec \tau^2 \vec C^2) =  \tau_j^2 \lambda_i^4 (\vec V_i \cdot \vec t_j)^2, 
\eal 
where summation over $i$ and $j$ is implied. Setting the above four terms to be constant can be seen as four independent equations for the $\vec V_i$'s and $\vec t_j$'s. Being unit eigenvectors of symmetric matrices, they must also satisfy the twelve equations
\be
\vec V_i\cdot \vec V_m = \delta_{im} \quad \text{and} \quad \vec t_j\cdot \vec t_n = \delta_{jn},
\en 
for $i,j,m, n =1,2,3$. 

So in total, the $\vec V_i$'s and $\vec t_j$'s have to satisfy 16 equations for the 18 unknown components of  $\vec V_1$, $\vec V_2$, $\vec V_3$, $\vec t_1$, $\vec t_2$ and $\vec t_3$. This leaves us with two degrees of freedom which we can carefully use to reduce Eq.~\eqref{eqn:tauCoaxialC}. For example, let $\vec V_1 \cdot \vec t_1 = \vec V_2 \cdot \vec t_1 =0$, which implies that $\vec t_1 = \pm \vec V_3$ and so $\vec t_1$ is an eigenvector of both $\vec C$ and $\vec \tau$. By taking the dot product of Eq.~\eqref{eqn:tauCoaxialC} with $\vec t_1$ on the left and right side we conclude that
\be
\alpha_{ijkmn}  \tau^{j+m}_1 \lambda^{2(i+k+n)}_3  =0.
\label{eqn:tau1lambda3}
\en
Now, in the same way, we could have used the degrees of freedom to choose $\vec V_1 \cdot \vec t_2 = \vec V_2 \cdot \vec t_2 =0$ and concluded that
\be
\alpha_{ijkmn}  \tau^{j+m}_2 \lambda^{2(i+k+n)}_3  =0.
\label{eqn:tau2lambda3}
\en
and similarily choosing $\vec V_1 \cdot \vec t_3 = \vec V_2 \cdot \vec t_3 =0$ we conclude that
\be
\alpha_{ijkmn}  \tau^{j+m}_3 \lambda^{2(i+k+n)}_3  =0.
\label{eqn:tau3lambda3}
\en
We can rewrite the three above equations by using the characteristic polynomial of $\lambda^2_q$ and of $\tau_p$, 
\be
\label{eqns:CharacteristicPoly}
\lambda_3^{6} = I_{1} \lambda_3^4 - I_{2} \lambda_3^2 + 1, 
\quad \tau_3^3 = I_{\tau1} \tau_3^2 - I_{\tau2} \tau_3 + I_{\tau3},
\en
%\bal
%&\lambda_3^{8} = (I_{1}^2 - I_{2} )\lambda_3^4+ (1- I_{1} I_{2}) \lambda_3^2 + I_{1}  , 
%\\
%&\lambda_3^{6} = I_{1} \lambda_3^4 - I_{2} \lambda_3^2 + 1, 
%\\
%& \lambda_3^{-2} =\lambda_3^{4}- I_{1} \lambda_3^2+ I_{2},
%\\
%&\lambda_3^{-4} =I_{2} \lambda_3^{4} +(1- I_{2} I_{1})\lambda_3^{2} -I_{1} + I_{2}^2,
%\\
%&\tau_3^3 = I_{\tau1} \tau_3^2 - I_{\tau2} \tau_3 + I_{\tau3},
%\\
%&\tau_3^4 = (I_{\tau1}^2 - I_{\tau2} ) \tau_3^2 + (I_{\tau3} - I_{\tau1}I_{\tau2})  \tau_3 + I_{\tau1} I_{\tau3},
%\eal
to replace powers of $\lambda_3$ higher than $4$ and lower than $0$ with $\lambda_p^2$, $\lambda_p^4$ and the invariants of $\vec C$, and the analogous for powers of $\tau_1$, $\tau_2$ and $\tau_3$. In doing so Eqs.~\eqref{eqn:tau1lambda3},~\eqref{eqn:tau2lambda3} and~\eqref{eqn:tau3lambda3} respectively become
\be
\beta_{ij} \lambda^{2 i}_3 \tau_1^j = 0, \;\; \beta_{ij} \lambda^{2 i}_3 \tau_2^j =0, \;\; \beta_{ij} \lambda^{2 i}_3 \tau_3^j =0,
\label{eqns:lambda3}
\en
for $i,j = 0,1$ and $2$, and where 
%\bal
%&\alpha_{i2k2n}  ((I_{\tau1}^2 - I_{\tau2} ) \tau_3^2 + (I_{\tau3} - I_{\tau1}) I_{\tau2} \tau_3 + I_{\tau1} I_{\tau3} ) \lambda^{2(i+k+n)}_3  =0.
%\\
%&\alpha_{ijkmn}  \tau^{j+m}_3 \lambda^{2(i+k+n)}_3  =0.
%\\
%&\alpha_{ijkmn}  \tau^{j+m}_3 \lambda^{2(i+k+n)}_3  =0.
%\eal
the $\beta_{ij}$'s can be written in terms of $\alpha_{ijkmn}$ and the invariants of $\vec \tau$ and $\vec C$. This is best done with a computer algebra system. 

As a reminder, the eigenvectors of $\vec C$ and $\vec \tau$ in each of the above three equations are most likely different. 
 However, the eigenvectors only appear in the form of the mixed invariants~\eqref{eqns:Icombined}, which are the same for the three equations because the of Eqs.~\eqref{eqns:ConstantMixedInvariants1} and~\eqref{eqns:ConstantMixedInvariants2}. Thus we may solve all the above three Eqs.~(\ref{eqn:tau1lambda3}), (\ref{eqn:tau2lambda3}) and (\ref{eqn:tau3lambda3}) simultaneously. 
 
 The three equations state that $\tau_1$, $\tau_2$ and $\tau_3$ are the roots of the second-order polynomial $\beta_{ij} \lambda^{2 i}_3 x^j =0$. As these eigenvalues are assumed to be all different, this is only possible if
\be
\beta_{ij} \lambda^{2 i}_3 =0, \quad \text{for } j=0,1,2.
\en  
Finally, there was nothing special about $\vec V_3$ and $\lambda_3$, using analogous arguments we can conclude that
\be
\beta_{ij} \lambda^{2 i}_1 =0, \;\; \beta_{ij} \lambda^{2 i}_2 = 0, \; \; \beta_{ij} \lambda^{2 i}_3 =0, \quad \text{for } j=0,1,2.
\en  
Again, $\lambda_1$,$\lambda_2$ and $\lambda_3$ are assumed to be all different, so the only way that $\lambda_1$, $\lambda_2$ and $\lambda_3$ are the roots of a second-order polynomial is if the polynomials coefficients are all zero, leading to
\be
\beta_{ij}  = 0, \quad \text{for } i,j=0,1,2.
\en
This result suggest a simple way to reach expression for the $\beta_{ij}$'s: we can substitute scalars $\lambda$ and $\tau$, in place of the tensors \eqref{eqn:tauCoaxialC}, followed by repeatedly using the characteristic polynomials of both $\lambda$ and $t$ as explained around Eq.~\eqref{eqns:CharacteristicPoly}, and then the $\beta_{ij}$'s will be the coefficients of the multivariate polynomial in $\lambda$ and $\tau$. The results are explicity given by Eqs.~\eqref{eqn:ISS1} and~\eqref{eqn:ISS2} combined with Appendix~\ref{app:ISSmatrices}. 
  
\section{ISS matrices}
\label{app:ISSmatrices}

The matrices appearing in Eqs.~\eqref{eqn:ISS1} and~\eqref{eqn:ISS2} have the following expressions:

\be
\vec P_{\{0\}}^{\{3\}}=
\Scale[\scalemath]{
-4
\begin{bmatrix}
 2\Psi_{I_1}^2 + 2\Psi_{I_2}(p -I_2 \Psi_{I_2}) \\
 p^2/2 -4 \Psi_{I_1}\Psi_{I_2} -2 I_1 \Psi_{I_2}^2 \\
 2 I_1 \Psi_{I_1}^2 + 2\Psi_{I_2}^2 - 2 \Psi_{I_1} (p -2 I_2 \Psi_{I_2}) \\
 0 \\
 0 \\
 0
\end{bmatrix}
}, \;
\vec P_{\{1\}}^{\{3\}}=
\Scale[\scalemath]{
8
\begin{bmatrix}
 0 \\
 0\\
 0\\
 -2 \Psi_{I1}\\
 2 I_2 \Psi_{I_1} +2 \Psi_{I_2} \\
 p -2 I_1 \Psi_{I_1} -2 I_2\Psi_{I_2} 
\end{bmatrix}
},
\en
\be
\vec P_{\{2\}}^{\{3\}}=
\Scale[\scalemath]{
16
\begin{bmatrix}
 0 \\
 0\\
 0\\
 p -2 I_1 \Psi_{I_1} -2 I_2 \Psi_{I_2}\\
 2(I_1 I_2 -1) \Psi_{I_1} +I_2(2 I_2 \Psi_{I_2} -p) \\
 I_1 p -2 (I_1^2 - I_2) \Psi_{I_1} +2(1 - I_1 I_2) \Psi_{I_2} 
\end{bmatrix}
},
\en
\be
\vec P_{\{3\}}^{\{3\}}=
\Scale[\scalemath]{
-8
\begin{bmatrix}
 I_{\tau3}(2 \Psi_{J_1} +4 I_1 \Psi_{J_2} + I_{\tau1} \Psi_{J3}) \\
 4 I_{\tau3} \Psi_{J_2}\\
 I_{\tau3} (2 I_1 \Psi_{J_1} +4 (I_1^2 -I_2)\Psi_{J_2} +I_1 I_{\tau1} \Psi_{J_3}) \\
 ( I_{\tau3}- I_{\tau1}I_{\tau2})\Psi_{J_3} -2 I_{\tau2} \Psi_{J_1}  -4 I_1 I_{\tau2} \Psi_{J_2}  
 \\
 2 I_2 I_{\tau2} \Psi_{J_1} +4(I_1 I_2 -1)I_{\tau2} \Psi_{J_2} +(I_{\tau1}I_{\tau2} -I_{\tau3})I_2 \Psi_{J_3}  
 \\
 4 ( I_2-I_1^2 )I_{\tau2}\Psi_{J_2} -2 I_1 I_{\tau2} \Psi_{J_1} +(I_{\tau3}- I_{\tau1}I_{\tau2})I_1\Psi_{J_3}
\end{bmatrix}
},
\en
\begin{multline}
\vec P^{\{3\}}_{\{4\}} =
\Scale[\scalemath]{
32
\begin{bmatrix}
 I_{\tau3}  \left(2 \left(I_2 -I_1^2\right) \Psi_{J_2}-I_1 \Psi_{J_1}\right) \\
 -I_{\tau3}  \left(2 I_1 \Psi_{J_2}+\Psi_{J_1}\right) \\
 I_{\tau3}  \left(2 \left( 2 I_1 I_2-I_1^3-1\right)
   \Psi_{J_2}+\left( I_2-I_1^2\right) \Psi_{J_1}\right) \\
 I_{\tau2} \left(2 \left(I_1^2-I_2\right) \Psi_{J_2}+I_1 \Psi_{J_1}\right) 
 \\
 I_{\tau2} \left(2 \left(  I_1 +I_2^2- I_1^2 I_2\right) \Psi_{J_2}+( 1-I_1 I_2) \Psi_{J_1}\right) 
 \\
 I_{\tau2} \left(2 \left(I_1^3-2 I_1 I_2+1\right)
   \Psi_{J_2}+\left(I_1^2-I_2\right) \Psi_{J_1}\right) 
\end{bmatrix}
} + 
\\
\Scale[\scalemath]{
32
\begin{bmatrix}
 -I_{\tau1} I_{\tau3}  \left(\left(I_1^2-I_2\right) \Psi_{J_4}+I_1
   \Psi_{J_3}\right) \\
 -I_{\tau1} I_{\tau3}  \left(I_1 \Psi_{J_4}+\Psi_{J_3}\right) \\
 -I_{\tau1} I_{\tau3}  \left(\left(I_1^3-2 I_1 I_2+1\right)
   \Psi_{J_4}+\left(I_1^2-I_2\right) \Psi_{J_3}\right) \\
 (I_{\tau1} I_{\tau2}-I_{\tau3} ) \left(\left(I_1^2-I_2\right)
   \Psi_{J_4}+I_1 \Psi_{J_3}\right) \\
 ( I_{\tau1} I_{\tau2}- I_{\tau3}) \left(\left( I_1+I_2^2 -I_1^2 I_2\right)
   \Psi_{J_4}+(1-I_1 I_2) \Psi_{J_3}\right) \\
 (I_{\tau1} I_{\tau2}-I_{\tau3} ) \left(\left(I_1^3-2 I_1 I_2+1\right)
   \Psi_{J_4}+\left(I_1^2-I_2\right) \Psi_{J_3}\right) 
\end{bmatrix}
},
\end{multline}
\be
\vec P^{\{4\}}_{\{0\}} =
\Scale[\scalemath]{
-4
\begin{bmatrix}
 -4 I_1 \Psi_{I_2}^2-8 \Psi_{I_1} \Psi_{I_2}+p^2 \\
 -4 I_1 p \Psi_{I_2}-4 p \Psi_{I_1}+4 \Psi_{I_2}^2+I_2 p^2 \\
 4 \left(\Psi_{I_1}^2+\Psi_{I_2} \left(p-I_2 \Psi_{I_2}\right)\right) \\
 0 \\
 0 \\
 0 \\
\end{bmatrix}
},
\en
\be
\vec P^{\{4\}}_{\{1\}} =
\Scale[\scalemath]{
16
\begin{bmatrix}
 0 \\
 0 \\
 0 \\
 2 \Psi_{I_2} \\
 p-2 I_2 \Psi_{I_2} \\
 -2 \Psi_{I_1} \\
\end{bmatrix}
}, \quad
\vec P^{\{4\}}_{\{2\}} =
\Scale[\scalemath]{
32
\begin{bmatrix}
 0 \\
 0 \\
 0 \\
 -2 \Psi_{I_1} \\
 2 \left(I_2 \Psi_{I_1}+\Psi_{I_2}\right) \\
 -2 I_1 \Psi_{I_1}-2 I_2 \Psi_{I_2}+p \\
\end{bmatrix}
},
\en
\be
\vec P^{\{4\}}_{\{3\}} =
\Scale[\scalemath]{
-16
\begin{bmatrix}
 4 I_{\tau3}  \Psi_{J_2} \\
 0 \\
 I_{\tau3}  \left(4 I_1 \Psi_{J_2}+I_{\tau1} \Psi_{J_3}+2 \Psi_{J_1}\right) \\
 -4 I_{\tau2} \Psi_{J_2} \\
 4 I_2 I_{\tau2} \Psi_{J_2} \\
 \Psi_{J_3} (I_{\tau3}- I_{\tau1} I_{\tau2} )-4 I_1 I_{\tau2} \Psi_{J_2}-2
   I_{\tau2} \Psi_{J_1} \\
\end{bmatrix}
},
\en
\be
\vec P^{\{4\}}_{\{4\}} =
\Scale[\scalemath]{
-64
\begin{bmatrix}
 I_{\tau3}  \left(2 I_1 \Psi_{J_2}+\Psi_{J_1}\right) \\
 2 I_{\tau3}  \Psi_{J_2} \\
 I_{\tau3}  \left(2 \left(I_1^2-I_2\right) \Psi_{J_2}+I_1 \Psi_{J_1}\right) \\
 -2 I_1 I_{\tau2} \Psi_{J_2}-I_{\tau2} \Psi_{J_1} \\
 I_{\tau2} ( 2  (I_1 I_2-1) \Psi_{J_2}+I_2  \Psi_{J_1}) \\
 I_{\tau2} (2 \left( I_2-I_1^2\right) \Psi_{J_2}-I_1 \Psi_{J_1} ) 
\end{bmatrix}
-64
\begin{bmatrix}
 I_{\tau1} I_{\tau3}  \left(I_1 \Psi_{J_4}+\Psi_{J_3}\right) \\
 I_{\tau1} I_{\tau3}  \Psi_{J_4} \\
 I_{\tau1} I_{\tau3}  \left(\left(I_1^2-I_2\right) \Psi_{J_4}+I_1
   \Psi_{J_3}\right) \\
 -(I_{\tau1} I_{\tau2}-I_{\tau3} ) \left(I_1 \Psi_{J_4}+\Psi_{J_3}\right) \\
 (I_{\tau1} I_{\tau2}-I_{\tau3} ) \left((I_1 I_2-1) \Psi_{J_4}+I_2
   \Psi_{J_3}\right) \\
 (I_{\tau1} I_{\tau2}-I_{\tau3} ) \left(\left(I_2 -I_1^2\right)
   \Psi_{J_4}-I_1 \Psi_{J_3}\right) 
\end{bmatrix}
},
\en
\be 
\vec Q_{\{1\}}^{\{1\}}= \vec Q_{\{2\}}^{\{1\}} = \vec Q_{\{1\}}^{\{2\}} = \vec Q_{\{2\}}^{\{2\}}= \vec Q_{\{0\}}^{\{n\}}  = \vec 0, \;\; \text{for} \; n=1,2,3 \; \text{and } \,4,
\en 
\be
\vec Q^{\{1\}}_{\{3\}} = 
\Scale[\scalemath]{
-4
\begin{bmatrix}
 0 \\
 0 \\
 1 \\
\end{bmatrix}
}, \;\;
\vec Q^{\{1\}}_{\{4\}} = 
\Scale[\scalemath]{
8
 \begin{bmatrix}
 -1 \\
 I_2 \\
 -I_1 \\
\end{bmatrix}
}
, \; \;
\vec Q_{\{3\}}^{\{2\}} =
\Scale[\scalemath]{
-8 
\begin{bmatrix}
 0 \\
 1 \\
 0 \\
\end{bmatrix}
},\;\;
\vec Q_{\{4\}}^{\{2\}} =
\Scale[\scalemath]{
-16 
\begin{bmatrix}
 0 \\
 0 \\
 1 \\
\end{bmatrix}
},
\en
\be
\vec Q_{\{1\}}^{\{3\}} =
\Scale[\scalemath]{
8 
\begin{bmatrix}
- \Psi_{J_1} \\
 I_2 \Psi_{J_1} \\
 -I_1 \Psi_{J_1} 
\end{bmatrix}
},\;\;
\vec Q_{\{2\}}^{\{3\}} =
\Scale[\scalemath]{
-32
 \begin{bmatrix}
 I_1 \Psi_{J_1}+\left(I_1^2-I_2\right) \Psi_{J_2} \\
 (1-I_1 I_2) \Psi_{J_1}+\left(-I_2 I_1^2+I_1+I_2^2\right) \Psi_{J_2} \\
 \left(I_1^2-I_2\right) \Psi_{J_1}+\left(I_1^3-2 I_2 I_1+1\right) \Psi_{J_2} 
\end{bmatrix}
},
\en
\be
\vec Q_{\{3\}}^{\{3\}} =
\Scale[\scalemath]{
8
 \begin{bmatrix}
 -\Psi_{J_3} I_{\tau1}^2-2 \Psi_{J_1} I_{\tau1}-4 I_1 \Psi_{J_2} I_{\tau1}-2 \Psi_{I_1}+I_{\tau2}
   \Psi_{J_3} \\
 I_2 \Psi_{J_3} I_{\tau1}^2+2 I_2 \Psi_{J_1} I_{\tau1}+4 I_1 I_2 \Psi_{J_2} I_{\tau1}-4 \Psi_{J_2}
   I_{\tau1}+2 I_2 \Psi_{I_1}+2 \Psi_{I_2}-I_2 I_{\tau2} \Psi_{J_3} \\
 I_{\tau2} \Psi_{J_3} I_1+p-2 I_2 \Psi_{I_2}+ 4 I_{\tau1} I_2  \Psi_{J_2}-I_{\tau1}I_1 \left( 4 \Psi_{J_2} I_1 +2 \Psi_{I_1}+2  \Psi_{J_1}  +I_{\tau1}  \Psi_{J_3} \right )
\end{bmatrix}
},
\en
\begin{multline}
\vec Q_{\{4\}}^{\{3\}} =
\Scale[\scalemath]{
32 
 \left(I_{\tau1}^2-I_{\tau2}\right) \Psi_{J_4} 
\begin{bmatrix}
 I_2-I_1^2 \\
 I_2 I_1^2-I_1-I_2^2 \\
 2 I_2 I_1-1-I_1^3 \\
\end{bmatrix}
 + 
32 \left(I_{\tau1}^2-I_{\tau2}\right) \Psi_{\text{J3}} \begin{bmatrix}
 -I_1 \\
 I_1 I_2-1 \\
 I_2-I_1^2 \\
\end{bmatrix}
+ } 
\\
\Scale[\scalemath]{
16 \begin{bmatrix}
 p-2 I_2 \Psi_{I_2}-2 I_1 \left(\Psi_{I_1}+I_{\tau1} \Psi_{J_1}\right)+4 \left(I_2-I_1^2\right) I_{\tau1}
   \Psi_{\text{J2}} \\
 2 (I_1 I_2-1) \Psi_{I_1}-I_2 p-2 I_{\tau1} \Psi_{J_1}+2 I_2 \left(I_2 \Psi_{I_2}+I_1 I_{\tau1}
   \Psi_{J_1}\right)-4 \left(-I_2 I_1^2+I_1+I_2^2\right) I_{\tau1} \Psi_{\text{J2}} \\
 2 \left(I_2-I_1^2\right) \Psi_{I_1}+I_1 p+2(1- I_1 I_2) \Psi_{I_2}+2 \left(I_2-I_1^2\right) I_{\tau1} \Psi_{J_1}-4 \left(I_1^3-2 I_2 I_1+1\right) I_{\tau1} \Psi_{\text{J2}} 
\end{bmatrix}
},
\end{multline}
\be
\vec Q_{\{1\}}^{\{4\}} =
\Scale[\scalemath]{
-16
 \begin{bmatrix}
 0 \\
 0 \\
 \Psi_{J_1} 
\end{bmatrix}
},\;\; 
\\
\vec Q_{\{2\}}^{\{4\}} =
\Scale[\scalemath]{
64
 \begin{bmatrix}
 -\Psi_{J_1}-I_1 \Psi_{J_2} \\
 I_2 \Psi_{J_1}+(I_1 I_2-1) \Psi_{J_2} \\
 \left(I_2-I_1^2\right) \Psi_{J_2}-I_1 \Psi_{J_1} \\
\end{bmatrix}
},
\en
\be
\vec Q_{\{3\}}^{\{4\}} =
\Scale[\scalemath]{
16
\begin{bmatrix}
 2 \left(\Psi_{I_2}-2 I_{\tau1} \Psi_{J_2}\right) \\
 p-2 I_2 \Psi_{I_2}+4 I_2 I_{\tau1} \Psi_{J_2} \\
 -\Psi_{J_3} I_{\tau1}^2-2 \Psi_{J_1} I_{\tau1}-4 I_1 \Psi_{J_2} I_{\tau1}-2 \Psi_{I_1}+I_{\tau2}
   \Psi_{J_3} 
\end{bmatrix}
},
\en
\begin{multline}
\vec Q_{\{4\}}^{\{4\}} =
\Scale[\scalemath]{
64 \left(I_{\tau1}^2-I_{\tau2}\right) \Psi_{J_3} 
\begin{bmatrix}
 -1 \\
 I_2 \\
 -I_1 \\
\end{bmatrix}
 +
64 \left(I_{\tau1}^2-I_{\tau2}\right) \Psi_{\text{J4}} 
\begin{bmatrix}
 -I_1 \\
 I_1 I_2-1 \\
 I_2-I_1^2 \\
\end{bmatrix} 
+}  
\\
\Scale[\scalemath]{
32 \begin{bmatrix}
 2 \left(I_2 \Psi_{I_1}+\Psi_{I_2}+I_2 I_{\tau1} \Psi_{J_1}+2 (I_1 I_2-1) I_{\tau1} \Psi_{\text{J2}}\right) \\
 -2 \left(\Psi_{I_1}+I_{\tau1} \left(\Psi_{J_1}+2 I_1 \Psi_{\text{J2}}\right)\right) \\
 p-2 I_2 \Psi_{I_2}-2 I_1 \left(\Psi_{I_1}+I_{\tau1} \Psi_{J_1}\right)+4 \left(I_2-I_1^2\right) I_{\tau1}
   \Psi_{\text{J2}} \\
\end{bmatrix}
},
%+64
%\Scale[\scalemath]{
 %\begin{bmatrix}
 %-\Psi_{I_1}-I_{\tau1} \Psi_{J_1}+I_1 \left(-\Psi_{J_4} I_{\tau1}^2-2 \Psi_{J_2} I_{\tau1}+I_{\tau2}
   %\Psi_{J_4}\right) \\
 %I_1 I_2 \Psi_{J_4} I_{\tau1}^2-\Psi_{J_4} I_{\tau1}^2+I_2 \Psi_{J_1} I_{\tau1}+2 I_1 I_2
   %\Psi_{J_2} I_{\tau1}-2 \Psi_{J_2} I_{\tau1}+I_2 \Psi_{I_1}+\Psi_{I_2}-I_1 I_2 I_{\tau2}
   %\Psi_{J_4}+I_{\tau2} \Psi_{J_4} \\
 %-2 I_{\tau1} \Psi_{J_2} I_1^2-I_{\tau1}^2 \Psi_{J_4} I_1^2+I_{\tau2} \Psi_{J_4} I_1^2-\Psi_{I_1}
   %I_1-I_{\tau1} \Psi_{J_1} I_1+\frac{p}{2}-I_2 \Psi_{I_2}+2 I_2 I_{\tau1} \Psi_{J_2}+I_2 I_{\tau1}^2 \Psi_{J_4}-I_2 I_{\tau2} \Psi_{J_4} \\
%\end{bmatrix} 
%} 
\end{multline}

\end{appendices}

\end{document}